\documentclass[aps,twocolumn,prc,showpacs]{revtex4}
\usepackage{mathrsfs}
\usepackage{longtable,lscape}
\usepackage{txfonts}
\usepackage{amssymb}
\usepackage{indentfirst}
\usepackage{graphicx,,booktabs}
\usepackage{color}
\usepackage{amssymb}
\usepackage{epsfig}

\newcommand{\vsig}{\mbox{\boldmath$\sigma$\unboldmath}}
\newcommand{\veps}{\mbox{\boldmath$\epsilon$\unboldmath}}

\newcommand{\be}{\begin{equation}}
\newcommand{\ee}{\end{equation}}
\newcommand{\bea}{\begin{eqnarray}}
\newcommand{\eea}{\end{eqnarray}}
\newcommand{\bean}{\begin{eqnarray*}}
\newcommand{\eean}{\end{eqnarray*}}

\newcommand{\gapproxeq}{\lower
.7ex\hbox{$\;\stackrel{\textstyle >}{\sim}\;$}}
\newcommand{\lapproxeq}{\lower
.7ex\hbox{$\;\stackrel{\textstyle <}{\sim}\;$}}

\begin{document}

\title{\textbf{ Charmonium spectrum and their electromagnetic transitions with higher multipole contributions}}
\author{
Wei-Jun Deng, Hui Liu, Long-Cheng Gui~\footnote {E-mail: guilongcheng@ihep.ac.cn}, and Xian-Hui Zhong
\footnote {E-mail: zhongxh@hunnu.edu.cn} } \affiliation{ 1) Department
of Physics, Hunan Normal University, and Key Laboratory of
Low-Dimensional Quantum Structures and Quantum Control of Ministry
of Education, Changsha 410081, China }

\affiliation{ 2) Synergetic Innovation
Center for Quantum Effects and Applications (SICQEA), Changsha 410081, China}

\begin{abstract}
The charmonium spectrum is calculated with two nonrelativistic quark models,
the linear potential model and the screened potential model.
Using the obtained wavefunctions, we evaluate the electromagnetic transitions of charmonium states
up to $4S$ multiplet. The higher multipole contributions are included by a
multipole expansion of the electromagnetic interactions. Our results
are in reasonable agreement with the measurements. As conventional charmonium states,
the radiative decay properties of the newly observed charmonium-like states,
such as $X(3823)$, $X(3872)$, $X(4140/4274)$, are discussed. The $X(3823)$ as $\psi_2(1D)$,
its radiative decay properties well agree with the observations.
From the radiative decay properties of $X(3872)$, one can not exclude it as a $\chi_{c1}(2P)$ dominant state.
We also give discussions of possibly observing the missing charmonium states in radiative
transitions, which might provide some useful references to look for them in forthcoming experiments.
The higher multipole contributions to the electromagnetic transitions are analyzed as well.
It is found that the higher contribution from the magnetic part could
give notable corrections to some E1 dominant processes by interfering with the E1 amplitudes.
Our predictions for the normalized magnetic quadrupole amplitudes $M_2$ of the $\chi_{c1,2}(1P)\to J/\psi \gamma$ processes
are in good agreement with the recent CLEO measurements.
\end{abstract}


\maketitle

\section{Introduction}

During the past few years, great progress has been achieved in the
observation of the charmonia~\cite{Eichten:2007qx,Brambilla:2010cs,Bevan:2014iga,Chen:2016qju,Voloshin:2007dx}.
From the review of the Particle Data Group (PDG)~\cite{PDG}, one can see that
many new charmonium-like $``XYZ"$ states above open-charm
thresholds states have been observed at Belle, BaBar, LHC, BESIII, CLEO and so on.
The observations of these new states not only deepen our understanding of the
charmonium physics, but also bring us many mysteries in this field
to be uncovered~\cite{Brambilla:2010cs,Voloshin:2007dx,Chen:2016qju}. If these
newly observed $``XYZ"$ states, such as $X(3872)$, $X(3915)$, $X(4140/4274)$ and $Y(4260)$,
are assigned as conventional charmonium states, some properties, such as measured mass and decay
modes may be inconsistent with the predictions. Thus,
how to identify these newly observed charmonium-like $``XYZ"$ states and
how to understand their uncommon nature are great challenges for physicist.

Stimulated by the extensive progress made in the observation of the
charmonia, in this work we study the mass spectrum and electromagnetic (EM)
transitions of charmonium within the widely used
linear potential model~\cite{Godfrey:1985xj,Swanson:2005,Godfrey:2015dia}, and
the screened potential model~\cite{Li:2012vc,Li:2009zu}. As we know,
the EM decays of a hadron are sensitive to its inner structure. The
study of the EM decays not only is crucial for us to determine the
quantum numbers of the newly observed charmonium states, but also
provides very useful references for our search for the missing
charmonium states in experiments. To study the charmonium spectrum
and/or their EM decays, beside the widely used potential
models~\cite{Eichten:1979ms,Godfrey:1985xj,Godfrey:2015dia,Eichten:2002qv,Swanson:2005,
Segovia:2008zz,ChaoKT93,ChaoKT90,Li:2012vc,Li:2009zu,Barnes:2003vb,Cao:2012du},
some other models, such as lattice
QCD~\cite{Dudek:2006ej,Dudek:2009kk,Becirevic:2014rda,Chen:2011kpa,Becirevic:2012dc,Yang:2012mya,Donald:2012ga,Liu:2012ze},
QCD sum rules~\cite{Khodjamirian:1979fa,Beilin:1984pf,Zhu:1998ih}, coupled-channel
quark models~\cite{Eichten:2004uh},
effective Lagrangian approach~\cite{DeFazio:2008xq,Wang:2011rt},
nonrelativistic effective field theories of
QCD~\cite{Brambilla:2005zw,Brambilla:2012be,Pineda:2013lta,Jia:2010jn},
relativistic quark model~\cite{Ebert:2003}, relativistic Salpeter
method~\cite{Wang:2010ej}, light front quark model~\cite{Ke:2013zs},
Coulomb gauge approach~\cite{Guo:2014zva}, and generalized
screened potential model~\cite{Gonzalez:2015hqa} have been employed in
theory. Recently, the hadronic loop contributions to the
radiative decay of charmonium states were also discussed
in Refs.~\cite{Li:2007xr,Zhao:2013jza,Li:2011ssa}.
Although there are many studies about the EM decays of charmonium states,
many properties are not well understood.
For example the predictions for the $\chi_{cJ}(1P)\to J/\psi \gamma$ and $\psi(3770)\to \chi_{cJ}(1P)\gamma$
processes are rather different in various models~\cite{BESIII:2015cby}. These
differences may come from the wavefunctions of charmonium states adopted,
the higher EM multipole amplitude contributions, the coupled-channel effects and so on.
Thus, to clarify these puzzles, more studies are needed.

In this work, we mainly focus on the following issues: (i) To clearly show the
model dependence of the higher charmonium states, we calculate
the charmonium spectroscopy within two typical models, i.e., linear and screened potential models.
As done in the literature, e.g.~\cite{Godfrey:1985xj,Swanson:2005,Godfrey:2015dia,Cao:2012du}, the spin-dependent
potentials are dealt with non-perturbatively so that
the corrections of the spin-dependent interactions to
the wavefunctions can be included. (ii) We further analyze the EM
transitions between charmonium states. Based on the
obtained radiative decay properties and mass spectrum, we discuss
the classifications of the newly observed charmonium-like states;
while for the missing excited states we suggest strategies to find them in radiative
transitions. (iii) Finally, we discuss the possible higher EM multipole
contributions to a EM transition process.

The paper is organized as follows. In Sec.~\ref{spec},
the charmonium spectroscopy is calculated
within both the linear and screened potential models.
In Sec.~\ref{EM}, firstly we give an introduction
of EM transitions described in present work. Then,
using the wavefunctions obtained from both the linear
and screened potential models, we analyze the EM decays of
charmonium states. Finally, a summary is
given in Sec.~\ref{sum}.

\begin{table}[htb]
\begin{center}
\caption{Charmonium mass spectrum. LP and SP stand for our calculated masses
with the linear potential and screened
potential models, respectively.  For comparison, the
measured masses (MeV) from the PDG~\cite{PDG}, and the previous
predictions with screened potential in Ref.~\cite{Li:2009zu} and linear potential
in Ref.~\cite{Swanson:2005} are also listed.  } \label{tab1s}
\begin{tabular}{ccccccccc}
\hline\hline
 $n^{2S+1}L_J$ & name &$J^{PC}$ &Exp.~\cite{PDG}  &~\cite{Swanson:2005} &~\cite{Li:2009zu}&LP &SP \\
 \hline
$1 ^3S_{1}$               &$J/\psi$        &$1^{--}$    &$3097\footnote{These masses for the 12
well-esbalished $c\bar{c}$ states are used as input to determine the model parameters.}$
&3090     &3097     &3097 &3097\\
$1 ^1S_{0}$         &$\eta_{c}(1S)$        &$0^{-+}$    &$2984^a$       &2982     &2979     &2983 &2984\\
$2 ^3S_{1}$             &$\psi(2S)$        &$1^{--}$    &$3686^a$       &3672     &3673     &3679 &3679\\
$2 ^1S_{0}$         &$\eta_{c}(2S)$        &$0^{-+}$    &$3639^a$       &3630     &3623     &3635 &3637\\
$3 ^3S_{1}$             &$\psi(3S)$        &$1^{--}$    &$4040^a$       &4072     &4022     &4078 &4030\\
$3 ^1S_{0}$         &$\eta_{c}(3S)$        &$0^{-+}$    &             &4043     &3991     &4048 &4004\\
$4 ^3S_{1}$             &$\psi(4S)$        &$1^{--}$    & 4415?       &4406     &4273     &4412 &4281\\
$4 ^1S_{0}$         &$\eta_{c}(4S)$        &$0^{-+}$    &             &4384     &4250     &4388 &4264\\
$5 ^3S_{1}$             &$\psi(5S)$        &$1^{--}$    &             &         &4463     &4711 &4472\\
$5 ^1S_{0}$         &$\eta_{c}(5S)$        &$0^{-+}$    &             &         &4446     &4690 &4459\\
$1 ^3P_{2}$        &$\chi_{c2}(1P)$        &$2^{++}$    &$3556^a$       &3556     &3554     &3552 &3553\\
$1 ^3P_{1}$        &$\chi_{c1}(1P)$        &$1^{++}$    &$3511^a$       &3505     &3510     &3516 &3521\\
$1 ^3P_{0}$        &$\chi_{c0}(1P)$        &$0^{++}$    &$3415^a$       &3424     &3433     &3415 &3415\\
$1 ^1P_{1}$            &$h_{c}(1P)$        &$1^{+-}$    &$3525^a$       &3516     &3519     &3522 &3526\\
$2 ^3P_{2}$        &$\chi_{c2}(2P)$        &$2^{++}$    &$3927^a$       &3972     &3937     &3967 &3937\\
$2 ^3P_{1}$        &$\chi_{c1}(2P)$        &$1^{++}$    &             &3925     &3901     &3937 &3914\\
$2 ^3P_{0}$        &$\chi_{c0}(2P)$        &$0^{++}$    &$3918?$      &3852     &3842     &3869 &3848\\
$2 ^1P_{1}$            &$h_{c}(2P)$        &$1^{+-}$    &             &3934     &3908     &3940 &3916\\
$3 ^3P_{2}$        &$\chi_{c2}(3P)$        &$2^{++}$    &             &4317     &4208     &4310 &4211\\
$3 ^3P_{1}$        &$\chi_{c1}(3P)$        &$1^{++}$    &             &4271     &4178     &4284 &4192\\
$3 ^3P_{0}$        &$\chi_{c0}(3P)$        &$0^{++}$    &             &4202     &4131     &4230 &4146\\
$3 ^1P_{1}$            &$h_{c}(3P)$        &$1^{+-}$    &             &4279     &4184     &4285 &4193\\
$1 ^3D_{3}$         &$\psi_{3}(1D)$        &$3^{--}$    &             &3806     &3799     &3811 &3808\\
$1 ^3D_{2}$         &$\psi_{2}(1D)$        &$2^{--}$    &$3823^a$       &3800     &3798     &3807 &3807\\
$1 ^3D_{1}$             &$\psi_1(1D)$        &$1^{--}$  &$3778^a$       &3785     &3787     &3787 &3792\\
$1 ^1D_{2}$        &$\eta_{c2}(1D)$        &$2^{-+}$    &             &3799     &3796     &3806 &3805\\
$2 ^3D_{3}$         &$\psi_{3}(2D)$        &$3^{--}$    &             &4167     &4103     &4172 &4112\\
$2 ^3D_{2}$         &$\psi_{2}(2D)$        &$2^{--}$    &             &4158     &4100     &4165 &4109\\
$2 ^3D_{1}$             &$\psi_1(2D)$        &$1^{--}$  &$4191?$       &4142     &4089     &4144 &4095\\
$2 ^1D_{2}$        &$\eta_{c2}(2D)$        &$2^{-+}$    &             &4158     &4099     &4164 &4108\\
$3 ^3D_{3}$         &$\psi_{3}(3D)$        &$3^{--}$    &             &         &4331     &4486 &4340\\
$3 ^3D_{2}$         &$\psi_{2}(3D)$        &$2^{--}$    &             &         &4327     &4478 &4337\\
$3 ^3D_{1}$             &$\psi_1(3D)$        &$1^{--}$  &             &         &4317     &4456 &4324\\
$3 ^1D_{2}$        &$\eta_{c2}(3D)$        &$2^{-+}$    &             &         &4326     &4478 &4336\\
\hline\hline
\end{tabular}
\end{center}
\end{table}

\begin{figure*}[ht] \centering \epsfxsize=5.86 cm \epsfbox{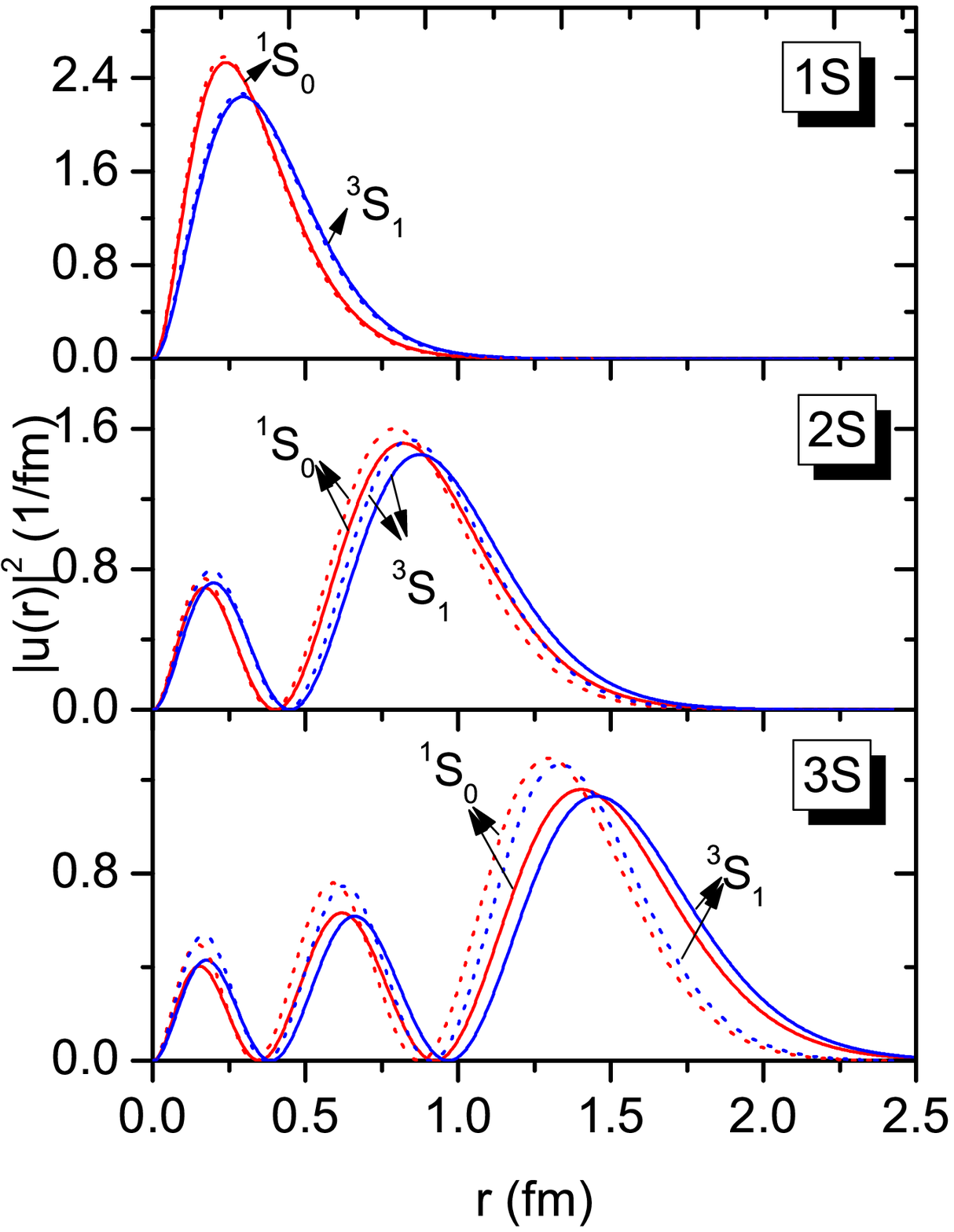}\epsfxsize=5.86 cm \epsfbox{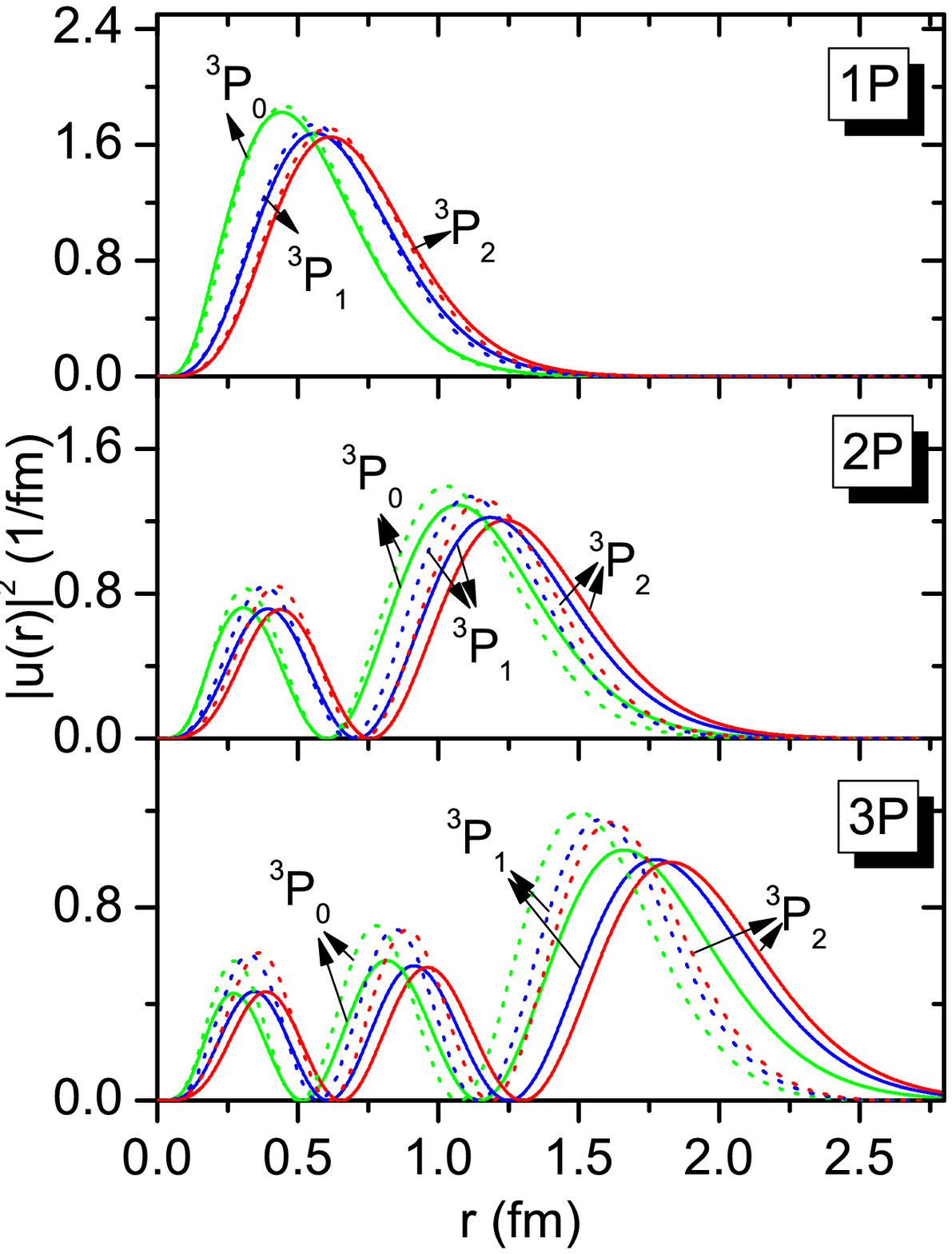}
\epsfxsize=5.86 cm \epsfbox{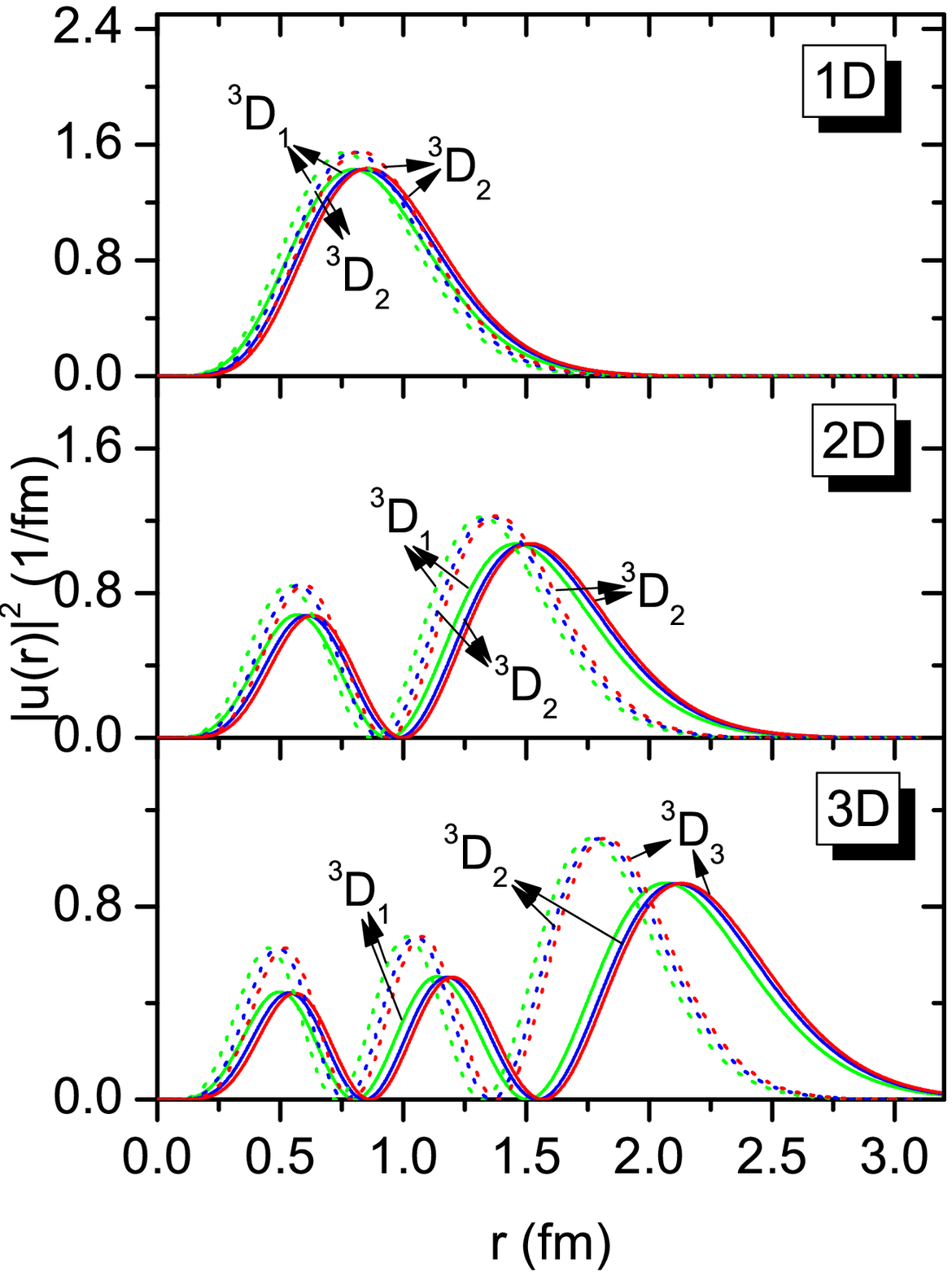}
\vspace{-0.5cm} \caption{(Color online) Predicted radial probability density
$|u(r)|^2$ for $S$-, $P$- and $D$-wave charmonium states up to $n=3$ shell.
The dotted and solid curves stand for the results obtained from the linear
and screened potential models, respectively.
}\label{wf}
\end{figure*}

\section{Mass spectroscopy} \label{spec}


\subsection{FORMALISM} \label{fa}

In this work, the mass and space wavefunction of a charmonium state are determined by the Schr\"{o}dinger
equation with a conventional quarkonium potential.
The effective potential of spin-independent term $V(r)$ between
the quark and antiquark is regarded as
the sum of Lorentz vector $V_V(r)$ and Lorentz scalar $V_s(r)$
contributions~\cite{Eichten:2007qx}, i.e.,
\begin{eqnarray}\label{vr}
V(r)=V_V(r)+V_s(r).
\end{eqnarray}
The Lorentz vector potential $V_V(r)$ is adopted the standard color
Coulomb form:
\begin{eqnarray}\label{vv}
V_V(r)=-\frac{4}{3}\frac{\alpha_s}{r}.
\end{eqnarray}
The Lorentz scalar $V_s(r)$ might be taken as
       \begin{eqnarray}
     V_s(r) =\cases{br,  & linear potential\cr
         \frac{b}{\mu}(1-e^{-\mu r}),  &screened potential},
 \end{eqnarray}
where $r$ is the distance between the quark and antiquark.
The linear potential $br$ is widely used in
the potential models. Considering the screening effect
from the vacuum polarization effect of the dynamical light quark
might soft the linear potential at large
distances~\cite{Laermann:1986pu,Born:1989iv}, people
suggested a screened potential $b(1-e^{-\mu r})/\mu$
in the calculations as well~\cite{Li:2009zu,Li:2012vc,Chao:2009,ChaoKT90,ChaoKT93}. Here $\mu$
is the screening factor which makes the long-range scalar potential
of $V_s(r)$ behave like $br$ when $r\ll 1/\mu$, and becomes a
constant $b/\mu$ when $r\gg 1/\mu$. The main effects of the screened
potential on the spectrum is that the masses of the higher excited
states are lowered.

Following the method in Refs.~\cite{Swanson:2005,Li:2009zu}, we include three spin-dependent
potentials in our calculations. For the
spin-spin contact hyperfine potential, we take the Gaussian-smeared form~\cite{Swanson:2005}
\begin{eqnarray}\label{ss}
H_{SS}= \frac{32\pi\alpha_s}{9m_c^2}\tilde{\delta}_\sigma(r)\mathbf{S}_c\cdot \mathbf{S}_{\bar{c}},
\end{eqnarray}
where $\mathbf{S}_c$ and $\mathbf{S}_{\bar{c}}$  are spin matrices acting on the spins
of the quark and antiquark. We take $\tilde{\delta}_\sigma(r)=(\sigma/\sqrt{\pi})^3
e^{-\sigma^2r^2}$ as in Ref.~\cite{Swanson:2005}.
The five parameters in the above equations ($\alpha_s$, $b$, $\mu$, $m_c$, $\sigma$)
are determined by fitting the spectrum.

For the spin-orbit term and the tensor term, we take the common
forms obtained from the leading-order perturbation theory:
\begin{eqnarray}\label{sl}
H_{SL}= \frac{1}{2m_c^2r}\left(3\frac{dV_V}{dr}-\frac{dV_s}{dr}\right)\mathbf{L}\cdot \mathbf{S},
\end{eqnarray}
and
\begin{eqnarray}\label{t}
H_{T}= \frac{1}{12m_c^2}\left(\frac{1}{r}\frac{dV_V}{dr}-\frac{d^2V_V}{dr^2}\right)S_T,
\end{eqnarray}
where $\mathbf{L}$ is the relative orbital angular momentum of $c$ and $\bar{c}$ quarks,
$\mathbf{S}=\mathbf{S}_c+\mathbf{S}_{\bar{c}}$ is the total quark spin, and the
spin tensor $S_T$ is defined by
\begin{eqnarray}\label{st}
S_T= 6\frac{\mathbf{S}\cdot \mathbf{r}\mathbf{S}\cdot \mathbf{r}}{r^2}-2\mathbf{S}^2.
\end{eqnarray}

By solving the radial Schr\"{o}dinger equation
$\frac{d^2u(r)}{dr^2}+2\mu_R \left[E-V_{c\bar{c}}(r)-\frac{L(L+1)}{2\mu_R r^2}\right]u(r)=0$,
with $V_{c\bar{c}}(r)\equiv V(r)+H_{SS}+H_{SL}+H_{T}$
and $\mu_R\equiv m_c m_{\bar{c}}/(m_c+m_{\bar{c}})$, we obtain the wavefunction $u(r)$ and 
the mass $M_{c\bar{c}}=2m_c+E$ for a charmonium state.
For simplification, the spin-dependent
interactions can be dealt with perturbatively. Although the meson mass
obtains perturbative corrections from these spin-dependent
potentials, the wave functions obtain no corrections from them.
Thus, to reasonably include the corrections
from these spin-dependent potentials to both the mass and wave
function of a meson state, we deal with the spin-dependent
interactions nonperturbatively.

In this work, we solve the radial Schr\"{o}dinger equation by using
the three-point difference central method from central ($r=0$)
towards outside ($r\to \infty$) point by point. The details of this
method can be found in Ref.~\cite{Haicai}. To overcome the singular
behavior of $1/r^3$ in the spin-dependent potentials, following
the method of our previous work~\cite{Deng:2016ktl}, we introduce a cutoff distance
$r_c$ in the calculation. Within a small range $r\in (0,r_c)$, we let
$1/r^3=1/r_c^3$. It is found that the masses of the $^3P_0$ states are sensitive to
the cutoff distance $r_c$, which is easily determined by the mass
of $\chi_{c0}(1P)$.

Considering the progress in the charmonium spectrum in recent years,
we do not use the old parameter sets determined in Refs.~\cite{Li:2009zu,Swanson:2005}.
Combining the new measurements, we slightly adjust the parameter sets
of Refs.~\cite{Li:2009zu,Swanson:2005} to better describe the data. By fitting the masses of the 12
well-established $c\bar{c}$ states given in Tab.~\ref{tab1s}, we obtain
the parameter sets for the linear potential model and screened potential model,
which are given in Tab.~\ref{qkp}.

\begin{table}[htb]
\begin{center}
\caption{ Quark model parameters determined by the 12
well-established $c\bar{c}$ states given in Tab.~\ref{tab1s}. }\label{qkp}
\begin{tabular}{c|c|c}
\hline\hline
  Parameter    & Linear potential model & Screened potential model \\
 \hline
    $m_c$  (GeV)         &  1.4830  &   1.4110                     \\
    $\alpha_s$           &  0.5461  &   0.5070                     \\
    $b$      (GeV$^2$)   &  0.1425  &   0.2100                     \\
    $\sigma$ (GeV)       &  1.1384  &   1.1600                     \\
    $r_c$    (fm)        &  0.202    &   0.180                       \\
    $\mu$ (GeV)          &  ...     &   0.0979                     \\
\hline\hline
\end{tabular}
\end{center}
\end{table}

\subsection{Results and discussions} \label{fa}

Our calculated masses for the $nS$ ($n\leq 5$), $nP$ and $nD$ ($n\leq 3$) charmonium states with both
the linear and screened potential models have been listed in Tab.~\ref{tab1s}, respectively.
It is found the mass spectrum calculated from the three-point difference central method
is consistent with the previous calculations~\cite{Li:2009zu,Swanson:2005}.
For the states with a mass of $M< 4.1$ GeV,
both linear and screened potential models
give a reasonable description of the mass spectrum compared with the data.
However, for the higher resonances with a mass of $M>4.1$ GeV,
the predictions between these two models are very different.
In the linear potential model, the
well-established states $\psi(4160)$ and $\psi(4415)$ could be
assigned as the $\psi_1(2D)$ and $\psi(4S)$, respectively.
However, in the screened potential model, the $\psi(4415)$
might be assigned as $\psi(5S)$~\cite{Li:2009zu}, while for
$\psi(4160)$, the predicted mass are about 100 MeV less than the measurements.
Comparing with the linear potential model,
an obvious feature of the screened potential model is that it provides a compressed mass spectrum,
which permits many new charmonium-like $``XYZ"$ states be accomodated
in the conventional higher charmonium states~\cite{Li:2009zu}.
Lately, BESIII Collaboration observed two resonant structures, one with a mass of $\sim 4222$ MeV and a width of $\sim 44$ MeV,
and the other with a mass of $\sim 4320$ MeV and a width of $\sim 101$ MeV,
in the cross section for the process $e^+e^-\to \pi^+\pi^-J/\psi $~\cite{Ablikim:2016qzw}, which may
correspond to the $J^{PC}$=$1^{--}$ states $X(4260)$ and $X(4360)$ from the PDG~\cite{PDG}, respectively.
It is found that within the screened potential model, $X(4260)$ and $X(4360)$
are good candidates of the $\psi(4S)$ and $\psi_1(3D)$, respectively.
Furthermore, we should mention that recently two new charmonium-like states
$X(4140)$ and $X(4274)$ were confirmed by the LHCb collaboration~\cite{Aaij:2016iza}.
Their quantum numbers are determined to be $J^{PC}=1^{++}$. Within the linear potential model the $X(4274)$ might
be identified as the $\chi_{c1}(3P)$ state. While within the screened potential model,
the $X(4140)$ is a good candidate of $\chi_{c1}(3P)$.
However, neither the linear potential model nor the screened model can give two
conventional $J^{PC}=1^{++}$ charmonium states with masses around 4.14 and 4.27 GeV
at the same time, which may indicate the exotic nature of $X(4140)$ and/or $X(4274)$.

Furthermore, in Tab.~\ref{tabhf}, we give our predictions of the
hyperfine splittings for some $S$-wave states, and fine splittings
for some $P$-wave states with the linear
and screened potentials, respectively. For a comparison, the world average data
from the PDG~\cite{PDG} and the previous predictions in
Refs.~\cite{Li:2009zu,Swanson:2005} are listed in the
same table as well. It is found that both the linear
and screened potential models give comparable results.
The predicted splittings are in agreement with the world average data~\cite{PDG}.
It should be mentioned that both the linear
and screened potential models obtain a similar fine splitting between $\chi_{c2}(2P)$ and
$\chi_{c0}(2P)$, i.e., $\Delta m\approx 90$ MeV. According to the measured mass of $\chi_{c2}(2P)$, one can
predict that the mass of $\chi_{c0}(2P)$ is about 3837 MeV. Thus, assigning the $X(3915)$ as the
$\chi_{c0}(2P)$ state is still problematic, which was also pointed out in
Refs.~\cite{Guo:2012tv,Olsen:2014maa,Zhou:2015uva}.

To better understand the properties of the wavefunctions of the charmonium states
which are important to the decays, we plot
the radial probability density as a function of the interquark
distance $r$ in Fig.~\ref{wf}. It is found that the spin-dependent potentials have
notable corrections to the $S$- and triplet $P$-wave states. The spin-spin potential $H_{SS}$
brings an obvious splitting to the wavefunctions between $n^1S_0$ and $n^3S_1$,
while the tensor potential $H_{T}$ brings notable
splittings to the wavefunctions between the triplet $P$-wave states.
The spin-dependent potentials only give a tiny correction to
wavefunctions of the higher triplet $nD$, $nF$, ... states.  On the other hand, comparing the results
from the linear potential model with those from the screened potential model,
we find that for the low-lying $1S,2S$ $1P$ and $1D$
charmonium states the wavefuctions obtained from both of the models are less different.
However, for the higher charmonium states $nS$ ($n\geq 3$),
$nP,nD...$ ($n\geq 2$), the wavefuctions
obtained from these two models show a notable difference.

\begin{table}[htb]
\begin{center}
\caption{ Hyperfine and fine splittings in units of MeV for charmonia. LP and SP stand our results
obtained from the linear potential and screened potential models, respectively.
The experimental data are taken from the PDG~\cite{PDG}. The theoretical
predictions with the previous screened potential model~\cite{Li:2009zu},
GI model and nonrelativistic linear potential model~\cite{Swanson:2005},
are also listed for comparison. }\label{tabhf}
\begin{tabular}{c|cccccccc}
\hline\hline
 Splitting & LP & SP &SNR~\cite{Li:2009zu} &NR~\cite{Swanson:2005} &GI~\cite{Swanson:2005} &Exp.~\cite{PDG} \\
 \hline
$m(1 ^3S_{1})$-$m(1 ^1S_{0})$    &  114 &   113      &118    &108      &113          &$113.3\pm 0.7$ \\
$m(2 ^3S_{1})$-$m(2 ^1S_{0})$    &  44 &    43       &50     &42       &53           &$46.7\pm 1.3$ \\
$m(3 ^3S_{1})$-$m(3 ^1S_{0})$    &  30 &    26       &31     &29       &36            &  \\
$m(4 ^3S_{1})$-$m(4 ^1S_{0})$    &  22 &    17       &       &22       &25            & \\
$m(5 ^3S_{1})$-$m(5 ^1S_{0})$    &  21 &    13       &       &         &            & \\

$m(1 ^3P_{2})$-$m(1 ^3P_{1})$    &  36 &   32        &44    &51       &40          &$45.5\pm 0.2$ \\
$m(1 ^3P_{1})$-$m(1 ^3P_{0})$     & 101 &  106        &77    &81       &65          &$95.9\pm 0.4$ \\
$m(2 ^3P_{2})$-$m(2 ^3P_{1})$     & 30 &  23         &36    &47       &26         &  \\
$m(2 ^3P_{1})$-$m(2 ^3P_{0})$     & 68 &  66         &59    &53       &37          & \\
$m(3 ^3P_{2})$-$m(3 ^3P_{1})$     & 25 &  19         &30    &46       &20          &\\
$m(3 ^3P_{1})$-$m(3 ^3P_{0})$     & 51 &  46         &47    &69       &25            & \\
\hline\hline
\end{tabular}
\end{center}
\end{table}

\section{electromagnetic transitions with higher multipole contributions}\label{EM}

Using the wavefunctions obtained from both the linear
and screened potential models, we further study the EM transitions
between charmonium states with higher multipole contributions.
The EM decay properties not only are crucial for us to determine the
quantum numbers of the newly observed charmonium states, but also
provide very useful references for our search for the missing
charmonium states in experiments.

\subsection{The model}\label{fb}

The quark-photon EM coupling at the tree level is
described by
\begin{eqnarray}\label{Hee}
H_e=-\sum_j
e_{j}\bar{\psi}_j\gamma^{j}_{\mu}A^{\mu}(\mathbf{k},\mathbf{r})\psi_j,
\end{eqnarray}
where $\psi_j$ stands for the $j$-th quark field in a hadron. The
photon has three momentum $\mathbf{k}$, and the constituent quark
$\psi_j$ carries a charge $e_j$.

In this work, the wave functions are calculated nonrelativistically from
the potential models. To match the nonrelativistic wave functions of
hadrons, we should adopt the nonrelativistic form of Eq.~(\ref{Hee}) in the
calculations. Including the effects of binding potential between quarks~\cite{Brodsky:1968ea},
the nonrelativistic expansion of $H_e$ may be written as~\cite{Close:1970kt,Li:1994cy,Li:1997gda}
\begin{equation}\label{he2}
h_{e}\simeq\sum_{j}\left[e_{j}\mathbf{r}_{j}\cdot\veps-\frac{e_{j}}{2m_{j}
}\vsig_{j}\cdot(\veps\times\hat{\mathbf{k}})\right]e^{-i\mathbf{k}\cdot
\mathbf{r}_{j}},
\end{equation}
where $m_j$, $\vsig_j$, and $\mathbf{r}_j$ stand for the constituent mass,
Pauli spin vector, and coordinate for the $j$-th quark, respectively.
The vector $\veps$ is the polarization vector of the photon.
It is found that the first and second terms in
Eq.(\ref{he2}) are responsible for the electric and
magnetic transitions, respectively. The second term
in Eq.(\ref{he2}) is the same as that used in
Ref.~\cite{Godfrey:1985xj}, while the first term in Eq.(\ref{he2})
differs from $(1/m_j) \mathbf{p}_j\cdot\veps$ used in Ref.~\cite{Godfrey:1985xj}
for the effects of the binding potential is included in the transition
operator. This nonrelativistic EM transition operator has been widely applied
to meson photoproduction reactions~\cite{Li:1995si,Li:1997gda,
Zhao:2001kk, Saghai:2001yd, Zhao:2002id,Zhao:1998fn,Zhao:1998rt, He:2008ty, He:2008uf,
He:2010ii, Zhong:2011ti, Zhong:2011ht, Xiao:2015gra}.

Finally the standard helicity transition amplitude $\mathcal{A}_\lambda$ between the initial
state $|J \lambda \rangle$ and final state $|J' \lambda' \rangle$ can be calculated by
\begin{eqnarray}\label{amp1}
\mathcal{A}_\lambda&=&-i\sqrt{\frac{\omega_\gamma}{2}}\langle J' \lambda' | h_{e}|J \lambda
\rangle,
\end{eqnarray}
where $\omega_\gamma$ is the photon energy.
It is easily found that the helicity amplitudes for the electric and magnetic
operators are
\begin{eqnarray}\label{amp2}
\mathcal{A}^{E}_\lambda&=&-i\sqrt{\frac{\omega_\gamma}{2}}\Big \langle J' \lambda'
\Big | \sum_{j}e_{j}\mathbf{r}_{j}\cdot\veps e^{-i\mathbf{k}\cdot
\mathbf{r}_{j}} \Big |J \lambda \Big \rangle,\\
\mathcal{A}^{M}_\lambda&=&+i\sqrt{\frac{\omega_\gamma}{2}}\Big \langle J' \lambda'
\Big | \sum_{j}\frac{e_{j}}{2m_{j}
}\vsig_{j}\cdot(\veps\times\hat{\mathbf{k}})e^{-i\mathbf{k}\cdot
\mathbf{r}_{j}}\Big | J \lambda\Big \rangle.
\end{eqnarray}

In the initial-hadron-rest system for the radiative decay precess,
the momentum of the initial hadron is $\bf{P}_i=0$, and that of the
final hadron state is $\bf{P}_f=-\textbf{k}$. Without losing generals,
we select the photon momentum along the $z$ axial ($\mathbf{k}=k\hat{\mathbf{z}}$), and
take the polarization vector of the photon with the right-hand  form, i.e.,
$\veps=-\frac{1}{\sqrt{2}}(1,i,0) $, in our calculations.
To easily work out the EM transition matrix elements,
we use the multipole expansion of the plane wave
\begin{eqnarray}\label{amp3}
e^{-i\mathbf{k}\cdot\mathbf{r}_{j}}=e^{-ikz_j}=\sum_l \sqrt{4\pi (2l+1)}(-i)^lj_l(kr_j)Y_{l0}(\Omega),
\end{eqnarray}
where $j_l(x)$ is the Bessel function, and $Y_{lm}(\Omega)$ are the well-known spherical harmonics.
Then, we obtain the matrix element for the electric multipole transitions
with angular momentum $l$ (E$l$ transitions)~\cite{Deng:2015bva}:
\begin{eqnarray}\label{amp4}
\mathcal{A}^{\mathrm{E}l}_\lambda &=&\sqrt{\frac{\omega_\gamma}{2}}\Big \langle J' \lambda'
\Big | \sum_{j} (-i)^{l}B_l e_{j}j_{l+1}(kr_j)r_{j} Y_{l1}  \Big | J \lambda \Big \rangle \nonumber\\
&+ &\sqrt{\frac{\omega_\gamma}{2}}\Big \langle J' \lambda'
\Big | \sum_{j} (-i)^{l}B_l e_{j}j_{l-1}(kr_j)r_{j} Y_{l1}  \Big | J \lambda \Big \rangle,
\end{eqnarray}
where $B_l\equiv\sqrt{\frac{2\pi l(l+1)}{2l+1}}$.
We also obtain the matrix element from the magnetic part
with angular momentum $l$ (M$l$ transitions):
\begin{eqnarray}\label{amp5}
\mathcal{A}^{\mathrm{M}l}_\lambda=\sqrt{\frac{\omega_\gamma}{2}}\Big \langle J' \lambda'
\Big | \sum_{j} (-i)^{l}C_l \frac{e_{j}}{2m_j} j_{l-1}(kr_j)\sigma^+_jY_{l-1~0}  \Big | J \lambda \Big \rangle\nonumber\\
=\sqrt{\frac{\omega_\gamma}{2}}\Big \langle J' \lambda'
\Big | \sum_{j} (-i)^{l}C_l \frac{e_{j}}{2m_j} j_{l-1}(kr_j)[\sigma^+_j\otimes Y_{l-1~0}]^l_1  \Big | J \lambda \Big \rangle\nonumber\\
+\sqrt{\frac{\omega_\gamma}{2}}\Big \langle J' \lambda'
\Big | \sum_{j} (-i)^{l}C_l \frac{e_{j}}{2m_j} j_{l-1}(kr_j)[\sigma^+_j\otimes Y_{l-1~0}]^{l-1}_1  \Big | J \lambda \Big \rangle,
\end{eqnarray}
where $C_l\equiv i\sqrt{8\pi (2l-1)}$, and $\sigma^+=\frac{1}{2}(\sigma_x+i\sigma_y)$ is the spin shift operator.
Obviously, the E$l$ transitions satisfy the parity selection rule: $\pi_i\pi_f=(-1)^l$;
while the M$l$ transitions satisfy the parity selection rule: $\pi_i\pi_f=(-1)^{l+1}$,
where $\pi_i$ and $\pi_f$ stand for the parities of the initial and final hadron states, respectively.
Finally, using the parity selection rules, one can express the EM helicity
amplitude $\mathcal{A}$ with the matrix elements of EM multipole transitions in a unified form:
\begin{eqnarray}\label{ampaa}
\mathcal{A}_\lambda= \sum_{l}\Big\{ \frac{1+(-1)^{\pi_i\pi_f+l}}{2}\mathcal{A}^{\mathrm{E}l}_\lambda
+ \frac{1-(-1)^{\pi_i\pi_f+l}}{2}\mathcal{A}^{\mathrm{M}l}_\lambda\Big\}.
\end{eqnarray}
Combining the parity selection rules, we easily know the possible
EM multipole contributions to a EM transition considered in present work,
which are listed in Tab.~\ref{tabs}.

It should be pointed out that, the second term of Eq.~(\ref{amp5}) from the
magnetic part is included into the electric part by the
most general decomposition of the helicity amplitudes~\cite{Karl:1975jp,Durand:1962zza,Olsson:1986aw}:
\begin{eqnarray}\label{Hdec}
\mathcal{A}_\lambda= \sum_{k\geq 1}(-1)^{k+1}\sqrt{\frac{2k+1}{2J+1}}a_k\langle k-1; J'\lambda+1| J\lambda\rangle,
\end{eqnarray}
with $ a_k$ corresponding to the multipole amplitude of the EM tensor operators with a rank $k$. The second term of Eq.~(\ref{amp5})
is called as ``extra" electric-multipole term, $E_R$, by F. E. Close \emph{et al.}
~\cite{Close:2002sf}. Specifically, for $^3S_1\leftrightarrow ^3P_1$:
\begin{eqnarray}\label{ass}
a_1&=& E_1+E_R= -\frac{\sqrt{2}}{2}(\mathcal{A}_0+\mathcal{A}_{-1}),\nonumber\\
a_2&=& M_2= -\frac{\sqrt{2}}{2}(\mathcal{A}_0-\mathcal{A}_{-1});
\end{eqnarray}
for $^3P_2\rightarrow ^3S_1$:
\begin{eqnarray}\label{ass}
a_1=E_1+E_R=\frac{\sqrt{10}}{2}(\sqrt{3}\mathcal{A}_{-1}-\mathcal{A}_0),\nonumber\\
a_2=M_2=\frac{\sqrt{6}}{2}(\sqrt{3}\mathcal{A}_0-\mathcal{A}_{-1});
\end{eqnarray}
and for $^3S_1\rightarrow ^3P_2$:
\begin{eqnarray}\label{ass}
a_1=E_1+E_R=\frac{\sqrt{10}}{2}(\sqrt{3}\mathcal{A}_0-\mathcal{A}_{-1}),\nonumber\\
a_2=M_2=\frac{\sqrt{6}}{2}(\mathcal{A}_0-\sqrt{3}\mathcal{A}_{-1}).
\end{eqnarray}
Here $E_1$ is the leading electric-dipole term determined by Eq.(\ref{amp4}),
and $M_2$ is the magnetic-quadrupole term related to
the first term of Eq.(\ref{amp5}). It should be mentioned that
we have $a_3=0$ for the above transitions in present work.

\begin{table}[htb]
\begin{center}
\caption{Possible
EM multipole contributions to a EM transition between two charmonium states. } \label{tabs}
\begin{tabular}{cccccccc}
\hline\hline
 process & multipole contribution  \\
 \hline
$n ^3S_{1}\longleftrightarrow m ^1S_{0}$               &M1  \\
$n ^3P_{J}\longleftrightarrow m ^3S_{1}$               &E1, M2  \\
$n ^1P_{1}\longleftrightarrow m ^1S_{0}$               &E1  \\
$n ^3D_{J}\longleftrightarrow m ^3P_{J}$               &E1, E3, M2, M4  \\
$n ^1D_{1}\longleftrightarrow m ^1P_{1}$               &E1, E3  \\
$n ^3P_{J}\longleftrightarrow m ^1P_{1}$               &M1, M3  \\
\hline\hline
\end{tabular}
\end{center}
\end{table}

Then, the partial decay widths of the EM transitions are given by
\begin{eqnarray}\label{dww}
\Gamma &=&\frac{|\mathbf{k}|^2}{\pi}\frac{2}{2J_i+1}\frac{M_{f}}{M_{i}}\sum_{\lambda}|\mathcal{A}_{\lambda}|^2\nonumber\\
       &=&\frac{|\mathbf{k}|^2}{\pi}\frac{2}{2J_i+1}\frac{M_{f}}{M_{i}}\sum_{k}|a_{k}|^2,
\end{eqnarray}
where $J_i$ is the total angular momenta of the initial mesons,
$J_{fz}$ and $J_{iz}$ are the components of the total angular
momentum along the $z$ axis of initial and final mesons,
respectively. To take into account of the relativistic effects, following the idea of Ref.~\cite{Swanson:2005},
we introduce an overall relativistic phase space factor $E_f/M_i$ in our predictions of the widths, which is
usually not far from unity. $M_f$ and $E_f$ stand for the mass and total energy of the final charmonium state, respectively.
$M_i$ is the mass of the initial charmonium state.

Finally, we should mention that in the most general decomposition of the
helicity amplitudes~\cite{Karl:1975jp,Durand:1962zza,Olsson:1986aw}, the $a_k$
is considered as the magnetic or electric multipole amplitude, thus,
in the total decay width the electric and magnetic multipole
amplitudes can not interfere with each other. However,
the ``extra" electric-multipole term comes from the magnetic part, thus, in this sense
the electric-magnetic interference term appears in the total decay width~\cite{Close:2002sf,Close:2002ky,Bonnaz:2001aj}.

\subsection{Results and discussions}\label{RD}

\subsubsection{ Lighter states }


First, we calculate the M1 transitions of the low-lying $1S$, $2S$ and $3S$ states. Our results
compared with experimental data and other model predictions have been listed in Tab.~\ref{EM2}.
Both our linear and screened potential model calculations obtain a compatible prediction.
Our predictions are consistent with those of NR and GI models~\cite{Swanson:2005}.
It should be pointed out that our predictions together with those in the framework of GI
and NR potential models~\cite{Swanson:2005} give a very
large partial width for the $\psi(2S)\to \eta_{c}(1S)\gamma$ process,
which is about an order of magnitude larger than the world average data
from the PDG~\cite{PDG} and the prediction of
$\Gamma[\psi(2S)\to \eta_{c}(1S)\gamma] \simeq 0.4(8)$ keV from
Lattice QCD~\cite{Dudek:2009kk}. Although our prediction of $\Gamma[J/\psi\rightarrow \eta_c(1S) \gamma]$
is obviously larger than the PDG average data~\cite{PDG}, it is in agreement with the recent measurement
$\Gamma[J/\psi\rightarrow \eta_c(1S) \gamma]\simeq 2.98\pm
0.18^{+0.15}_{-0.33}$ keV at KEDR~\cite{Anashin:2014wva}. As a whole,
strong model dependence exists in the predictions of the M1 transitions, more studies
are needed in both theory and experiments.

Then, we calculate the E1 dominant radiative decays of the $1P$ and $2S$
states. Our results compared with experimental data and other model predictions
have been listed in Tab.~\ref{EMPW1}. Both our linear and screened potential
model calculations obtain a compatible prediction, because the wave-functions and masses for
the low-lying states from these two models have less differences. Our predictions
are reasonable agreement with the data. The predictions from different models are
consistent with each other in a magnitude, although there are differences more or less.

\subsubsection{$\psi(3770)$, $X(3823)$ and the missing $1D$ states}


The $\psi(3770)$ resonance is primarily a $\psi_1(1D)$ state with
small admixtures of $\psi(2S)$~\cite{Eichten:2007qx}. It can decay
into $\chi_{cJ}(1P)\gamma$. These decay processes are dominated by the E1
transition. The radiative decays of $\psi(3770)$ are still not well understood.
For example, the predictions of $\Gamma[\psi(3770)\to \chi_{c0}(1P)\gamma]$ vary in
a very large range (200,500) keV~\cite{BESIII:2015cby}.
Considering $\psi(3770)$ as a pure $\psi_1(1D)$ state, we calculate the radiative decay widths of
$\Gamma[\psi(3770)\to \chi_{cJ}(1P)\gamma]$ with the wavefunctions obtained
from the linear and screened potential models, respectively.
Our results have been listed in Tab.~\ref{D-wave}. From the table, we can see
that both models give very similar predictions for the partial decay widths.
Considering the leading E$1$ decays only, our predictions are in agreement
with the world average data within their uncertainties~\cite{PDG}. However, including
the magnetic part, the partial decay widths predicted by us
are about a factor of $1.5$ larger than the world average data~\cite{PDG}
and the recent measurements from BESIII~\cite{BESIII:2015cby,Ablikim:2015sol}.
It is unclear whether these discrepancies are caused by our model limitations
or come from the experimental uncertainties. It should be mentioned that although some
predictions from the models with a relativistic assumption~\cite{Swanson:2005,Li:2009zu}
or a coupled-channel correction~\cite{Eichten:2004uh} seem to better
agree quantitatively with the experimental data, however,
the corrections of the magnetic part are not included in their calculations. To better understand the
radiative decay properties of $\psi(3770)$, more studies are needed
in both theory and experiments.


Recently, $X(3823)$ as a good candidate of $\psi_2(1D)$ was
observed by the Belle Collaboration in the $B\to \chi_{c1}\gamma K$
decay with a statistical significance of
$3.8\sigma$~\cite{Bhardwaj:2013rmw}. Lately, this state was
confirmed by the BESIII Collaboration in the process $e^+e^-\to
\pi^+\pi^-X(3823)\to \pi^+\pi^-\chi_{c1}\gamma$ with a statistical
significance of $6.2\sigma$~\cite{Ablikim:2015dlj}.
Assigning $X(3823)$ as the $\psi_2(1D)$ state, we predict the radiative decay
widths of $\Gamma[X(3823)\to \chi_{cJ}(1P)\gamma]$. Both the linear and
screened potential models give quite similar predictions,
\begin{eqnarray}
\Gamma[X(3823)\to \chi_{c1}(1P)\gamma]& \simeq & 300 \ \mathrm{keV},\\
\Gamma[X(3823)\to \chi_{c2}(1P)\gamma]& \simeq & 90 \ \mathrm{keV}.
\end{eqnarray}
Our prediction of $\Gamma[X(3823)\to \chi_{c1}(1P)\gamma]$
is close to the predictions in Refs.~\cite{Swanson:2005,Li:2009zu,Ebert:2003,Qiao:1996ve},
while our prediction for $\Gamma[X(3823)\to
\chi_{c2}(1P)\gamma]$ is about a factor of 1.4$\sim$1.8 larger than the predictions
in these works. Furthermore, our predicted partial width ratio,
\begin{eqnarray}
\frac{\Gamma[X(3823)\to \chi_{c2}(1P)\gamma]}{\Gamma[X(3823)\to
\chi_{c1}(1P)\gamma]}& \simeq &30\%,
\end{eqnarray}
is consistent with the observations $< 42\%$~\cite{Ablikim:2015dlj}.
The $X(3823)$ state mainly decays into the $\chi_{c1,2}(1P)\gamma$, $J/\psi \pi\pi$ and $ggg$ channels.
The predicted partial widths for $J/\psi \pi\pi$ and $ggg$ channels
are about $(210\pm 110)$ keV and 80 keV, respectively~\cite{Barnes:2003vb}. Thus, the total width
of the $X(3823)$ might be $\Gamma_{\mathrm{tot}}\simeq 680\pm 110$ keV, from which we obtain large
branching ratios
\begin{eqnarray}
Br[X(3823)\to \chi_{c1}(1P)\gamma]& \simeq &42\%,\\
Br[X(3823)\to \chi_{c2}(1P)\gamma]& \simeq &13\%.
\end{eqnarray}
The large branching fraction $Br[X(3823)\to \chi_{c1}(1P)\gamma]$
can explain why the $X(3823)$ was first observed in the $\chi_{c1}\gamma$ channel.


Another two $1D$-wave states $\psi_3(1D)$ and $\eta_{c2}(1D)$ have
not been observed in experiments. According to the theoretical
predictions, their masses are very similar to that of $\psi_2(1D)$.
If $X(3823)$ corresponds to the $\psi_2(1D)$ state indeed, the
masses of the $\psi_3(1D)$ and $\eta_{c2}(1D)$ resonances should be
around $3.82$ GeV. For the singlet $1D$ state $\eta_{c2}(1D)$, its main radiative
transition is $\eta_{c2}(1D)\to h_c(1P) \gamma$. This process
is governed by the E1 transition, the effects from the E3
transition are negligibly small. Taking the mass of
$\eta_{c2}(1D)$ with $M=3820$ MeV, with the wavefunctions
calculated from the linear potential model we predict that
\begin{eqnarray}
\Gamma[\eta_{c2}(1D)\to h_c(1P) \gamma]& \simeq & 362 \
\mathrm{keV},
\end{eqnarray}
which is consistent with that of the screened potential model. Our results are close to
the previous predictions in Refs.~\cite{Swanson:2005,Li:2009zu}
(see Tab.~\ref{D-wave}). Combined the predicted partial widths of
the other two main decay modes $gg$ and $\eta_c \pi\pi$~\cite{Barnes:2003vb},
the total width of $\eta_{c2}(1D)$ is estimated to be $\Gamma_{\mathrm{tot}}\simeq 760$ keV. Then,
we can obtain a large branching ratio
\begin{eqnarray}
Br[\eta_{c2}(1D)\to h_c(1P) \gamma]& \simeq &48\%.
\end{eqnarray}
Combining the measured branching ratios of $Br[ h_c(1P)\to \eta_c \gamma]\simeq 51\%$
and $Br[\eta_c\to K\bar{K}\pi]\simeq 7.3\%$~\cite{PDG}, we obtain
\begin{eqnarray}
Br[\eta_{c2}(1D)\to h_c(1P) \gamma\to \eta_c \gamma \gamma \to K\bar{K}\pi \gamma \gamma ] \simeq 1.8\%.
\end{eqnarray}
It should be mentioned that the $\eta_{c2}(1D)$ state could be produced via the $B\to \eta_{c2}(1D) K$ process
as suggested in Refs.~\cite{Fan:2009cj,Eichten:2002qv,Xu:2016kbn}.
The expectations are to accumulate $10^{10}$ $\Upsilon(4S)$ $B\bar{B}$ events at Belle~\cite{Godfrey:2015dia,Bevan:2014iga}.
If the branching fraction $Br[B\to \eta_{c2}(1D) K]$ is $\mathcal{O}(10^{-5})$ as predicted in~\cite{Xu:2016kbn},
the $10^{10}$ $B\bar{B}$ events could let us observe
$\mathcal{O}(1000)$ $\eta_{c2}(1D)$ events via the two-photon cascade
$\eta_{c2}(1D)\to h_c(1P) \gamma\to \eta_c \gamma \gamma$ in the
$\gamma\gamma K\bar{K}\pi$ final states.

While for the triplet $1D$ state $\psi_3(1D)$, its common radiative
transition is $\psi_3(1D)\to \chi_{c2}(1P)\gamma$. Taking the mass of
$\psi_3(1D)$ with $M=3830$ MeV, we calculate the partial decay
widths $\Gamma[\psi_{3}(1D)\to \chi_{c2}(1P) \gamma]$ with both the linear
and screened potential models. Both of the models give a very similar result
\begin{eqnarray}
\Gamma[\psi_{3}(1D)\to \chi_{c2}(1P)\gamma]& \simeq & 350 \
\mathrm{keV}.
\end{eqnarray}
The magnitude of the partial decay width of $\Gamma[\psi_{3}(1D)\to \chi_{c2}(1P)
\gamma]$ predicted by us is compatible with that in Refs.~\cite{Swanson:2005,Li:2009zu,Ebert:2003}.
Combining the predicted total width $\Gamma_{\mathrm{tot}}\simeq 3$ MeV for $\psi_3(1D)$~\cite{Barnes:2003vb}, we estimate
the branching ratio
\begin{eqnarray}
Br[\psi_{3}(1D)\to \chi_{c2}(1P)\gamma] \simeq 12\%.
\end{eqnarray}
The missing $\psi_{3}(1D)$ might be produced via the
$B\to \psi_3(1D) K$ process at Belle~\cite{Eichten:2002qv,Xu:2016kbn,Sang:2015lra},
and reconstructed in the $\chi_{c2}(1P) \gamma$ decay mode with $\chi_{c2}(1P)\to J/\psi \gamma$
and $J/\psi\to \mu^+\mu^-/e^+e^-$. If the branching fraction $Br[B\to \psi_{3}(1D)K]$
is $\mathcal{O}(10^{-5})$~\cite{Xu:2016kbn,Sang:2015lra},
based on the $10^{10}$ $B\bar{B}$ data to be accumulated at Belle II,
we expect that $\mathcal{O}(100)$ $\psi_{3}(1D)$ events could reconstructed
in the $\chi_{c2}(1P) \gamma$ channel with $\chi_{c2}(1P)\to J/\psi \gamma$
and $J/\psi\to \mu^+\mu^-/e^+e^-$.

\subsubsection{$X(3872,3915)$ and the $2P$ states  } 


In the $2P$-wave states, only $\chi_{c2}(2P)$ has been
established experimentally. This state was observed by both
Belle~\cite{Uehara:2005qd} and BaBar~\cite{Aubert:2010ab} in the
two-photon fusion process $\gamma\gamma\to D\bar{D}$ with a mass
$M\simeq 3927$ MeV and a narrow width $\Gamma\simeq 24$
MeV~\cite{PDG}. We analyze its radiative transitions to $\psi(1
D)\gamma$, $J/\psi\gamma$ and $\psi(2S)\gamma$.
Both the linear and screened potential give very similar predictions:
\begin{eqnarray}
\Gamma[\chi_{c2}(2P)\to \psi(3770)\gamma]& \simeq & 0.4 \
\mathrm{keV},\\
\Gamma[\chi_{c2}(2P)\to \psi_2(1D)\gamma]& \simeq & 3.2 \
\mathrm{keV},\\
\Gamma[\chi_{c2}(2P)\to \psi_3(1D)\gamma]& \simeq & 20 \
\mathrm{keV}.
\end{eqnarray}
Our predictions are notablely different from those of the NR
potential model~\cite{Swanson:2005} (see Tab.~\ref{EMPA}).
With the measured width, we further predicted the branching ratios
\begin{eqnarray}
Br[\chi_{c2}(2P)\to \psi(3770)\gamma]& \simeq & 1.7\times 10^{-5},\\
Br[\chi_{c2}(2P)\to \psi_2(1D)\gamma]& \simeq & 1.3\times 10^{-4},\\
Br[\chi_{c2}(2P)\to \psi_3(1D)\gamma]& \simeq & 1.5\times 10^{-3} .
\end{eqnarray}
Combining these ratios with the decay properties of $\psi(1D)$ and $\chi_c(1P)$ states,
we estimate the combined branching ratios for the decay chains
$\chi_{c2}(2P)\to \psi(1D)\gamma \to \chi_c(1P) \gamma\gamma \to J/\psi \gamma\gamma\gamma$,
our results have been listed in Tab.~\ref{2p2chain}. It is found that
the most important decay chains involving $\psi_2(1D)$ and $\psi_3(1D)$ are
$\chi_{c2}(2P)\to \psi_2(1D) \gamma \to \chi_{c1}(1P)\gamma\gamma\to J/\psi \gamma\gamma\gamma$ ($Br\simeq 1.9\times 10^{-5}$)
and $\chi_{c2}(2P)\to \psi_3(1D) \gamma\to \chi_{c2}(1P)\gamma\gamma\to J/\psi \gamma\gamma\gamma$ ($Br\simeq 1.4\times 10^{-5}$).
These decay chains might be hard observed at present because
the very small production cross section of $\chi_{c2}(2P)$.

The $\chi_{c2}(2P)$ state has relatively larger radiative decay rates
into $J/\psi\gamma$ and $\psi(2S)\gamma$. With the linear potential model, we obtain
\begin{eqnarray}
\Gamma[\chi_{c2}(2P)\to J/\psi\gamma]& \simeq & 93 \
\mathrm{keV},\\
\Gamma[\chi_{c2}(2P)\to \psi(2S)\gamma]& \simeq & 135 \
\mathrm{keV},
\end{eqnarray}
which are consistent with those of the screened potential model.
Combined the measured width, the branching ratios are predicted to be
\begin{eqnarray}
Br[\chi_{c2}(2P)\to J/\psi\gamma]& \simeq & 3.9\times 10^{-3},\\
Br[\chi_{c2}(2P)\to \psi(2S)\gamma]& \simeq & 5.6\times 10^{-3}.
\end{eqnarray}
It is might be a challenge to observe $\chi_{c2}(2P)$ in the $J/\psi\gamma$
and $\psi(2S)\gamma$ channels with $J/\psi/\psi(2S)\to \mu^+\mu^-$ at BESIII.
For example, we produce $\chi_{c2}(2P)$ via the process
$e^+e^-\to \gamma \chi_{c2}(2P)$ at BESIII. Based on the cross section
$\sim 0.2$ pb predicted in Ref.~\cite{Chao:2013cca}, we estimate that even if
BESIII can collect 10 fb$^{-1}$ data sample above open charm threshold,
we only accumulate about 2000 $e^+e^-\to \gamma \chi_{c2}(2P)$ events.
Combining our predictions of branching ratios $Br[\chi_{c2}(2P)\to J/\psi
\gamma\to \gamma\mu^+\mu^- ]\sim \mathcal{O}(10^{-4})$, we find
there is less hope for observing the radiative decay modes of
$\chi_{c2}(2P)$ at BESIII. It should be mentioned that the
decay chain $\chi_{c2}(2P)\to J/\psi\gamma,\psi(2S)\gamma\to \gamma\mu^+\mu^-$ might
be observed at Belle II or LHCb via the $B\to \chi_{c2}(2P) X$ decay.
The expectations are to accumulate $10^{10}$ $\Upsilon(4S)$ $B\bar{B}$
events at Belle~\cite{Godfrey:2015dia,Bevan:2014iga}. If the
branching ratio of $Br[B\to \chi_{c2}(2P) X]\sim 10^{-5}$,
we may observed $\mathcal{O}(10)$ $\chi_{c2}(2P)$'s
in the decay chain $\chi_{c2}(2P)\to J/\psi\gamma\to \gamma\mu^+\mu^- $.


The $\chi_{c1}(2P)$ state is still not established in experiments.
According to the fine splitting between $\chi_{c2}(2P)$ and $\chi_{c1}(2P)$,
we estimate the mass of $\chi_{c1}(2P)$ is around $M=3900$ MeV.
With this mass we calculate the transitions of $\chi_{c1}(2P)$
into $\psi(2S)\gamma$, $J/\psi \gamma$, $\psi_1(1D) \gamma$, and $\psi_2(1D) \gamma$.
Our results are listed in Tab.~\ref{EMPA}. With the wavefuctions
obtained from the linear potential model, it is found that the partial widths
\begin{eqnarray}
\Gamma[\chi_{c1}(2P)\to J/\psi\gamma]& \simeq & 81 \
\mathrm{keV},\\
\Gamma[\chi_{c1}(2P)\to \psi(2S)\gamma]& \simeq & 139 \
\mathrm{keV},
\end{eqnarray}
are slightly smaller than those from the screened potential model.
Combined the predicted width in Ref.~\cite{Swanson:2005},
the branching ratios might be
\begin{eqnarray}
Br[\chi_{c1}(2P)\to J/\psi\gamma]& \simeq & 4.9\times 10^{-4},\\
Br[\chi_{c1}(2P)\to \psi(2S)\gamma]& \simeq & 8.4\times 10^{-4}.
\end{eqnarray}
The partial widths of $\Gamma[\chi_{c1}(2P)\to \psi_{1,2}(1D) \gamma ]$ are about several keV,
their branching ratios are $\mathcal{O}(10^{-5})$.

The $X(3872)$ resonance has the same quantum numbers as
$\chi_{c1}(2P)$ (i.e., $J^{PC}=1^{++}$) and a similar mass to the
predicted value of $\chi_{c1}(2P)$. However, its exotic properties
can not be well understood with a pure $\chi_{c1}(2P)$
state~\cite{Olsen:2014qna,Voloshin:2007dx}. To understand the nature
of $X(3872)$, measurements of the radiative decays of $X(3872)$ have
been carried out by the BaBar~\cite{Aubert:2008ae},
Belle~\cite{Bhardwaj:2011dj}, and LHCb~\cite{Aaij:2014ala}
collaborations, respectively. Obvious evidence of $X(3872)\to J/\psi
\gamma$ was observed by these collaborations. Furthermore, the BaBar
and LHCb Collaborations also observed evidence of $X(3872)\to
\psi(2S) \gamma$. The branching fraction ratio
\begin{eqnarray}
R^{\mathrm{exp}}_{\psi'\gamma/\psi\gamma}=\frac{\Gamma[X(3872)\to
\psi(2S) \gamma]}{\Gamma[X(3872)\to J/\psi \gamma]}\simeq 3.4\pm
1.4,
\end{eqnarray}
obtained by the BaBar Collaboration~\cite{Aubert:2008ae} is
consistent with the recent measurement
$R^{\mathrm{exp}}_{\psi'\gamma/\psi\gamma}=2.46\pm 0.93$ of LHCb
Collaboration~\cite{Aaij:2014ala}.

Considering $X(3872)$ as a pure $\chi_{c1}(2P)$ state, we
calculate the radiative decays $X(3872)\to J/\psi \gamma,\psi(2S)
\gamma$. With the linear potential model, we predict that
\begin{eqnarray}
\Gamma[X(3872)\to J/\psi\gamma)& \simeq & 72 \
\mathrm{keV},\\
\Gamma[X(3872)\to \psi(2S)\gamma]& \simeq & 94 \ \mathrm{keV}.
\end{eqnarray}
With these predicted partial widths, we can easily obtain the ratio
\begin{eqnarray}
R^{\mathrm{th}}_{\psi'\gamma/\psi\gamma}=\frac{\Gamma[X(3872)\to
\psi(2S) \gamma]}{\Gamma[X(3872)\to J/\psi \gamma]}\simeq 1.3,
\end{eqnarray}
which is slightly smaller than the lower limit of
the measurements from the BaBar~\cite{Aubert:2008ae} and LHCb~\cite{Aaij:2014ala}.
Our predictions from the screened potential model are consistent
with those from the linear potential model.
Thus, from the view of branching fraction ratio
$R_{\psi'\gamma/\psi\gamma}$, we can not exclude the $X(3872)$
as a candidate of $\chi_{c1}(2P)$.

On the other hand, if $X(3872)$ corresponds to $\chi_{c1}(2P)$,
with the measured width (about several MeV)~\cite{PDG}, we can estimate
\begin{eqnarray}
Br[X(3872)\to J/\psi\gamma]&\sim & \mathcal{O}(10^{-2}),\\
Br[X(3872)\to \psi(2S)\gamma]&\sim & \mathcal{O}(10^{-2}).
\end{eqnarray}
Combining the branching ratio
$Br[B\to \chi_{c1}(2P) K]\sim O(10^{-4})$ predicted in Ref.~\cite{Meng:2005er},
we can further estimate that $Br[B\to \chi_{c1}(2P) K]\times Br[X(3872)
\to J/\psi\gamma/\psi(2S)\gamma]\sim O(10^{-6})$,
which is also consistent with the Belle measurements~\cite{Bhardwaj:2011dj}.


The $\chi_{c0}(2P)$ state is still not well-established, although
$X(3915)$ was recommended as the $\chi_{c0}(2P)$ state in
Ref.~\cite{Liu:2009fe}, and also assigned as the $\chi_{c0}(2P)$
state by the PDG recently~\cite{PDG}. Assigning $X(3915)$ as the
$\chi_{c0}(2P)$ state will face several serious
problems~\cite{Guo:2012tv,Olsen:2014maa}. Recently, Zhou \emph{et al.} carried out a
combined amplitude analysis of the $\gamma\gamma \to D\bar{D},
\omega J/\psi$ data~\cite{Zhou:2015uva}. They demonstrated that
$X(3915)$ and $X(3930)$ can be regarded as the same state with
$J^{PC}=2^{++}$ (i.e., $\chi_{c2}(2P)$). With the screened and linear potential models, our predicted masses for
the $\chi_{c0}(2P)$ state are $\sim3848$ MeV and $\sim3869$ MeV, respectively,
which are consistent with the previous predictions in
Refs.~\cite{Swanson:2005,Li:2009zu}, and the mass extracted by Guo and
Meissner by refitting the BaBar and Belle data of $\gamma\gamma \to
D\bar{D}$ separately~\cite{Guo:2012tv}. The $\chi_{c0}(2P)$ state can decay via the radiative transitions
$\chi_{c0}(2P)\to \psi(3770)\gamma, \psi(2S)\gamma, J/\psi \gamma$.
With the wavefunctions obtained from both the screened and linear potential models, we
calculate the decay rates of these radiative transitions. Our
results are listed in Tab.~\ref{EMPA}. From the table, it is found that both of
the models give similar predictions. The partial width for $\chi_{c0}(2P)\to \psi(2S)\gamma$ is
\begin{eqnarray}
\Gamma[\chi_{c0}(2P)\to \psi(2S)\gamma]& \simeq & 110\pm 10 \
\mathrm{keV}.
\end{eqnarray}
The $\chi_{c0}(2P)$ might be very broad with a width of $\sim 200$ MeV extracted from experimental data~\cite{Guo:2012tv},
which is about an order of magnitude larger than that predicted in Ref.~\cite{Swanson:2005}.
With the broad width, the branching ratio is predicted to be
\begin{eqnarray}
Br[\chi_{c0}(2P)\to \psi(2S)\gamma]& \simeq & 5.5\times 10^{-4}.
\end{eqnarray}
It should be mentioned that there is less chance for producing the $\chi_{c0}(2P)$ state via
the radiative decay chains $\psi(4040,4160,4415)\to \chi_{c0}(2P)\gamma\chi_{c0}(2P)
\to \psi(1S,2S) \gamma\gamma\to\gamma\gamma \mu^+\mu^-$ (see Tabs.~\ref{3stop}-\ref{4sp}).

There is no information of $h_{c}(2P)$ from experiments. According
to our predictions, the mass-splitting between $\chi_{c2}(2P)$
and $h_{c}(2P)$ is about $M_{\chi_{c2}(2P)}-M_{h_{c}(2P)}=(26\pm
4)$ MeV. Thus, the mass of
$h_{c}(2P)$ is most likely to be $M_{h_{c}(2P)}\simeq 3900$ MeV. The
typical radiative decay channels of $h_{c}(2P)$ are
$\eta_c(1S,2S)\gamma$ and $\eta_{c2}(1D) \gamma$. With the wavefunctions
obtained from the linear and screened potentials, we further calculate
these radiative decays. It is found that the radiative
transition rates of $h_{c}(2P)\to \eta_{c2}(1D) \gamma$,
$\eta_c(1S)\gamma$ and $\eta_c(2S)\gamma$ channels are fairly large.
Both the linear and screened potential models give very similar predictions
\begin{eqnarray}
\Gamma(h_{c}(2P)\to \eta_c(1S) \gamma)& \simeq &135 \
\mathrm{keV},\\
\Gamma(h_{c}(2P)\to \eta_c(2S)\gamma)& \simeq & 160 \
\mathrm{keV},\\
\Gamma(h_{c}(2P)\to \eta_{c2}(1D)\gamma)& \simeq & 25 \
\mathrm{keV}.
\end{eqnarray}
The rather sizeable partial widths for $h_{c}(2P)\to
\eta_c(1S,2S)\gamma$ are also obtained in the previous potential model
calculations~\cite{Swanson:2005,Li:2009zu,Cao:2012du}.
Combined the theoretical width $\Gamma\simeq 87$ MeV from~\cite{Swanson:2005},
the branching ratios are predicted to be
\begin{eqnarray}
Br[h_{c}(2P)\to \eta_c(1S) \gamma]& \simeq & 1.6\times 10^{-3},\\
Br[h_{c}(2P)\to \eta_c(2S) \gamma]& \simeq & 1.8\times 10^{-3},\\
Br[h_{c}(2P)\to \eta_{c2}(1D)\gamma]& \simeq & 2.8\times 10^{-4}.
\end{eqnarray}
The missing $h_{c}(2P)$ state might be produced via
$B\to h_{c}(2P) K$ process, and reconstructed in the
$\eta_c(1S,2S)\gamma$ decay modes with $\eta_c(1S,2S)\to K\bar{K}\pi$
at Belle II and LHCb.

\subsubsection{$\psi(4040)$ and the missing $\eta_c(3S)$ state}


The $\psi(4040)$ resonance is commonly identified with the
$\psi(3S)$ state~\cite{Eichten:2007qx}. This state can decay into
$\chi_{cJ}(1P)\gamma $ and $\chi_{cJ}(2P)\gamma $ via the radiative
transitions. We have calculated these precesses with both the linear
and screened potential model. Our results have been listed in Tab.~\ref{EMPW1}.
In our calculations, we find that the radiative transition rates of
$\psi(4040)\to \chi_{cJ}(1P)\gamma $ are relatively weak.
Using the PDG value for the total width $\Gamma\simeq 80$ MeV~\cite{PDG},
we obtain the branching ratios $Br[\psi(4040)\to \chi_{cJ}(1P)\gamma ]\sim \mathcal{O}(10^{-5})$,
which are consistent with the measurements $Br[\psi(4040)\to \chi_{c1,2}(1P)\gamma ]< 2\%$~\cite{PDG}.
Interestedly, it is found that the radiative transition rates of
$\psi(4040)\to \chi_{cJ}(2P)\gamma $ are rather sizeable. The
decay rates into the $\chi_{cJ}(2P)\gamma $ channels are about one
order of magnitude larger than those into the $\chi_{cJ}(1P)\gamma $
channels. With the screened potential model, we obtain that
\begin{eqnarray}
\Gamma[\psi(4040) \to \chi_{c2}(2P)\gamma]& \simeq & 82\
\mathrm{keV},\\
\Gamma[\psi(4040) \to \chi_{c1}(2P)\gamma]& \simeq & 67\
\mathrm{keV},\\
\Gamma[\psi(4040) \to \chi_{c0}(2P)\gamma]& \simeq & 27\
\mathrm{keV},
\end{eqnarray}
which are about $15\%$ larger than our linear potential model predictions.
Relatively large partial decay widths for $\psi(4040)\to \chi_{cJ}(2P)\gamma $ were
also found in the previous studies~\cite{Swanson:2005}. With the measured width
$\Gamma\simeq 80$ MeV from the PDG~\cite{PDG}, we
estimate the branching ratios
\begin{eqnarray}
Br[\psi(4040) \to \chi_{c2}(2P)\gamma]& \simeq & 1.0\times 10^{-3},\\
Br[\psi(4040) \to \chi_{c1}(2P)\gamma]& \simeq & 0.8\times 10^{-3},\\
Br[\psi(4040) \to \chi_{c0}(2P)\gamma]& \simeq & 3.3\times 10^{-4}.
\end{eqnarray}
It should be pointed out that BESIII plan to collect $5\sim 10$ fb$^{-1}$
$\psi(4040)$ in the coming years~\cite{XYZ2016}. Using the cross section of
$\sim 10$ nb based on BES and CLEO measurements~\cite{Bai:2001ct,Mo:2010bw,CroninHennessy:2008yi},
we expect to accumulate $(0.5\sim 1.0) \times 10^8$
$\psi(4040)$ events. The $\psi(4040)$ might provide us a source to produce
$\chi_{cJ}(2P)$ states via the radiative transitions. Thus,
we further estimate the number of events of the
two-photon cascades involving the $\chi_{cJ}(2P)$ states.
The results are listed in Tab.~\ref{3stop}. From the table, we can see that
about $\mathcal{O}(10)$ $\chi_{c2}(2P)$ events should be observed at BESIII via the radiative transition chain
$\psi(4040)\to \chi_{c2}(2P)\gamma \to J/\psi \gamma \gamma\to \gamma \gamma \mu^+\mu^-$.

The $\eta_c(3S)$ state is not established in experiments. According
to the model predictions, the hyperfine splitting between $3^3S_1$ and $3^1S_0$
is about 30 MeV (see Tab.~\ref{tabhf}). Thus, the mass of $\eta_c(3S)$ is
most likely to be $\sim 4010$ MeV. With this mass, we calculate the
radiative transitions $\eta_c(3S)\to h_c(1P)\gamma, h_c(2P)\gamma$.
Our prediction of the decay rate of $\eta_c(3S)\to h_c(1P)\gamma$ is tiny.
However, the partial decay widths are rather sizeable,
with the screened potential model we predict that
\begin{eqnarray}
\Gamma[\eta_c(3S) \to h_{c}(2P)\gamma]& \simeq & 130\ \mathrm{keV},
\end{eqnarray}
which is slightly ($\sim20\%$ ) larger than our prediction with the linear potential model.
Our prediction of $\Gamma[\eta_c(3S) \to h_{c}(2P)\gamma]$ is consistent with
the previous calculation in Ref.~\cite{Swanson:2005} (see Tab.~\ref{EM4s}).
Combined the predicted width $\Gamma\simeq 80$ MeV from~\cite{Swanson:2005}, the branching ratio of
$Br[\eta_c(3S) \to h_{c}(2P)\gamma]$ is estimated to be $1.6\times 10^{-3}$.

\subsubsection{$\psi(4160)$ and the missing $2D$ states}


The $1^{--}$ state $\psi(4160)$ is commonly identified with the $2^3
D_1$ state. The average experimental mass and width from the PDG are
$M=4191\pm 5$ MeV and $\Gamma=70\pm 10$ MeV,
respectively~\cite{PDG}, which are consistent with
linear potential model predictions. However, with
a screened potential, the predicted mass for $\psi_1(2D)$ is
about 100 MeV smaller than the observation.
The $\psi_1(2D)$ resonance can decay into
$\chi_{cJ}(1P)\gamma$ and $\chi_{cJ}(2P)\gamma$ via the radiative
transitions.

Considering $\psi(4160)$ as a pure $2^3 D_1$ state, with the
linear potential model, we predict that
\begin{eqnarray}
\Gamma[\psi(4160) \to \chi_{c0}(1P)\gamma]& \simeq & 150 \ \mathrm{keV},\\
\Gamma[\psi(4160) \to \chi_{c1}(1P)\gamma]& \simeq & 37\ \mathrm{keV},\\
\Gamma[\psi(4160) \to \chi_{c2}(1P)\gamma]& \simeq & 17\
\mathrm{keV}.
\end{eqnarray}
Similar results are also obtained with the screened potential model.
Our predictions of $\Gamma[\psi(4160) \to \chi_{c0,1}(1P)\gamma]$ are slightly
smaller than those obtained in Ref.~\cite{Segovia:2008zz}, however,
our predictions are notablely larger than those in Ref.~\cite{Swanson:2005} (see Tab.~\ref{D-wave}).
Combining the measured decay width of
$\psi(4160)$ with our predicted partial widths from the linear potential model, we estimate the branching fractions:
\begin{eqnarray}
Br[\psi(4160) \to \chi_{c0}(1P)\gamma]& \simeq & 2.1\times
10^{-3},\\
Br[\psi(4160) \to \chi_{c1}(1P)\gamma]& \simeq & 0.5\times
10^{-3},\\
Br[\psi(4160) \to \chi_{c2}(1P)\gamma]& \simeq & 0.2\times
10^{-3}.
\end{eqnarray}
Our predictions are in the range of the recent measurements
$Br[\psi(4160) \to \chi_{c1}(1P)\gamma]<6.1\times 10^{-3}$
and $Br[\psi(4160) \to \chi_{c2}(1P)\gamma]<16.2\times
10^{-3}$ from the Belle Collaboration~\cite{Han:2015vhc}.
We expect that more accurate observations can be carried out in future experiments.

Furthermore, we calculate the partial decay width of
$\Gamma[\psi(4160) \to \chi_{cJ}(2P)\gamma]$ with the linear and screened
potential models, respectively. Our results are listed in
Tab.~\ref{D-wave}. Both of the models give a similar result.
It is found that the decay rates of $\psi(4160)\to \chi_{c0}(2P)\gamma,\chi_{c1}(2P)\gamma$ are rater large,
their partial decay widths may be $300\sim 400$ keV. Similar results were also obtained
in Ref.~\cite{Swanson:2005,Li:2012vc}. The estimated branching ratios are
\begin{eqnarray}
Br[\psi(4160) \to \chi_{c2}(2P)\gamma]& \simeq & 0.3\times 10^{-3},\\
Br[\psi(4160) \to \chi_{c1}(2P)\gamma]& \simeq & 4.4\times 10^{-3},\\
Br[\psi(4160) \to \chi_{c0}(2P)\gamma]& \simeq & 4.4\times 10^{-3}.
\end{eqnarray}
It should be mentioned that $3$ fb$^{-1}$ new data of $\psi(4160)$
have been collected at BESIII~\cite{XYZ2016}.
Using the cross section of $\sim 8$ nb based on BES and CLEO measurements~\cite{Bai:2001ct,Mo:2010bw,CroninHennessy:2008yi},
we estimate that $2.4\times 10^{7}$ events of $\psi(4160)$ have been accumulated at BESIII.
Thus, if $\psi(4160)$ is the $2^3 D_1$ state indeed,
it might provide us a source to look for the missing $\chi_{c0}(2P)$ and
$\chi_{c1}(2P)$ states via the transition chains $\psi(4160)\to
\chi_{cJ}(2P)\gamma\to \psi(1S,2S)\gamma\gamma$.
The combined branching ratios of these decay chains, and the producing events
of $\chi_{cJ}(2P)$ estimated by us have been listed in Tab.~\ref{2d2p}.
It is found that if the $\chi_{cJ}(2P)$ is to be observed at BESIII via
the transition chains of $\psi(4160)\to
\chi_{cJ}(2P)\gamma\to \psi(1S,2S)\gamma\gamma\to\gamma\gamma \mu^+\mu^-$,
one should accumulate more data samples of $\psi(4160)$ in the coming years.

The other three $2D$-wave states, $\psi_2(2D)$, $\psi_3(2D)$ and $\eta_{c2}(2D)$, are
still not observed in experiments. With the masses and wavefuctions
predicted from the linear and screened potential models, we calculate
their radiative decay properties. Our results are listed in
Tab.~\ref{D-wave}. It is seen that although the predictions in details from
both linear and screened potential models have a notable difference, both models
predict that these $2D$ wave states $\psi_{2,3}(2D)$ and $\eta_{c2}(2D)$ have relatively large transition
rates into the $1P$- and $2P$-wave states. The partial
decay widths for the $2D\to 1P \gamma$ processes are about 10s keV
and their branching ratios are estimated to be $Br[2D\to 1P \gamma]\sim \mathcal{O}(10^{-4})$; while the partial
decay widths for the $2D\to 2P \gamma$ processes usually reach to 100s keV
and their branching ratios are estimated to be $Br[2D\to 2P \gamma]\sim \mathcal{O}(10^{-3})$.
The large decay rates of the $2D\to 2P \gamma$ processes were also predicted in
Ref.~\cite{Swanson:2005}. We further estimate the combined branching
ratios of the two-photon cascades $2D \to nP\to mS$.
Our results have been listed in Tab.~\ref{2d2p}.
In these decay chains, the most prominent two-photon cascades are $\psi_2(2D) \to \chi_{c1}(1P)\gamma
\to J/\psi \gamma\gamma\to \gamma\gamma \mu^+\mu^-$ ($Br\simeq 1.5\times 10^{-5}$), and $\eta_{c2}(2D)\to h_c(1P)
\gamma \to \eta_c \gamma\gamma\to \gamma\gamma K\bar{K}\pi$ ($Br\simeq 4.8\times 10^{-5}$). In coming years,
Belle II will accumulate $10^{10}$ $B\bar{B}$ data sample, which might let us obtain enough events of $2D$-wave states via $B\to \psi_{2}(2D)X$ and
$B\to \eta_{c2}(2D) X$ decays. If the branching fractions of $Br[B\to \psi_{2}(2D)X]$ and
$Br[B\to \eta_{c2}(2D) X]$ are $\mathcal{O}(10^{-5})$,
the missing $2D$-wave states might be observed in the above two-photon cascades.

\subsubsection{$X(4140,4274)$ and the $3P$ states}

Until now, no $3P$ charmonium states have been established in experiments.
According to the predicted masses and wave functions of the $3P$ charmonium states,
we estimate their radiative properties decay properties with both the linear and screened potential
models, which are listed in Tab.~\ref{EMPA}. From the table,
it is seen that most of our results from both models are similar in the magnitude.
The $\chi_{c0}(3P)$ state has a large decay rate into the $\psi(3S)\gamma$ channel,
the partial width might be 10s$\sim$100s keV, which is consistent with the
prediction in Ref.~\cite{Swanson:2005}. The $\chi_{c1,2}(3P)$/$h_{c}(3P)$ state has a large partial decay width
into $\psi(1S,2S,3S)\gamma$/$\eta_c(1S,2S,3S)\gamma$ channels, which are 10s$\sim$100s keV as well.
Combined the predicted widths from Ref.~\cite{Swanson:2005},
the estimated branching ratios of $Br[\chi_{c1,2}(3P)\to \psi(1S,2S,3S)\gamma]$ and
$Br[\chi_{c0}(3P)\to \psi(3S)\gamma]$ are $\mathcal{O}(10^{-3})$.
Using $B\to \chi_{c1,2}(3P)K/h_{c}(3P)K$ decays, the forthcoming Belle II and LHC experiments might
reconstruct these higher $\chi_{c1,2}(3P)$/$h_{c}(3P)$ states in
the $\psi(1S,2S)\gamma$/$\eta_c(2S)\gamma$ decay modes.

Recently, two new charmonium-like states
$X(4140)$ ($\Gamma\simeq 16$ MeV) and $X(4274)$ ($\Gamma\simeq 56$ MeV) are confirmed by the LHCb collaboration~\cite{Aaij:2016iza}.
Their quantum numbers are determined to be $J^{PC}=1^{++}$. According to the predicted
mass from the linear potential model, the $X(4274)$ might be a good candidate
of $\chi_{c1}(3P)$. However, within the screened potential model, $X(4140)$
seems to favor the $\chi_{c1}(3P)$ state. If the $X(4140)$ state is assigned as
$\chi_{c1}(3P)$, within the screened potential model the partial radiative decay widths of
the dominant channels are predicted to be
\begin{eqnarray}
\Gamma[\psi(4140) \to J/\psi\gamma]& \simeq & 38 \ \mathrm{keV},\\
\Gamma[\psi(4140) \to \psi(2S)\gamma]& \simeq & 51\ \mathrm{keV},\\
\Gamma[\psi(4140) \to \psi(3S)\gamma]& \simeq & 36\
\mathrm{keV},
\end{eqnarray}
Combined the average measured width with the predicted partial radiative decay widths of $X(4140)$,
the branching ratios are estimated to be
\begin{eqnarray}
Br[\psi(4140) \to J/\psi\gamma]& \simeq & 2.4\times 10^{-3} ,\\
Br[\psi(4140) \to \psi(2S)\gamma]& \simeq & 3.2\times 10^{-3},\\
Br[\psi(4140) \to \psi(3S)\gamma]& \simeq & 2.3\times 10^{-3}.
\end{eqnarray}
While, if the $X(4274)$ state is assigned as
$\chi_{c1}(3P)$, within the linear potential model the partial radiative decay widths of
the dominant channels are predicted to be
\begin{eqnarray}
\Gamma[\psi(4274) \to J/\psi\gamma]& \simeq & 48 \ \mathrm{keV},\\
\Gamma[\psi(4274) \to \psi(2S)\gamma]& \simeq & 88\ \mathrm{keV},\\
\Gamma[\psi(4274) \to \psi(3S)\gamma]& \simeq & 297\
\mathrm{keV},
\end{eqnarray}
Combined the measured width with the predicted partial radiative decay widths of $X(4274)$,
the branching ratios are estimated to be
\begin{eqnarray}
Br[\psi(4274) \to J/\psi\gamma]& \simeq & 0.9\times 10^{-3} ,\\
Br[\psi(4274) \to \psi(2S)\gamma]& \simeq & 1.6\times 10^{-3},\\
Br[\psi(4274) \to \psi(3S)\gamma]& \simeq & 5.3\times 10^{-3}.
\end{eqnarray}
The search for $X(4274)$ and $X(4140)$ in the $\psi(1S,2S,3S)\gamma$ channels and
the measurements of their partial width ratios might be helpful to uncover the nature
of these two newly observed states.

\subsubsection{$4S$ states}

In the $4S$ states, the $\psi(4S)$ resonance seems to favor the $1^{--}$ state
$\psi(4415)$ according to the linear potential model predictions~\cite{Swanson:2005}.
However, there are other explanations about $\psi(4415)$. For example in the screened potential model,
$\psi(4415)$ more favors $\psi(5S)$ other than $\psi(4S)$~\cite{Li:2009zu}, while with a
coupled-channel method the $\psi(4415)$ resonance are suggested to be
the $\psi_1(1D)$ resonance~\cite{Segovia:2008zz}. According to the screened potential model prediction,
the $J^{PC}=1^{--}$ state $X(4260)$ from the PDG~\cite{PDG}
could be a good candidate of the $\psi(4S)$. Very recently, BESIII Collaboration
observed a new structure $Y(4220)$ with a width of $\Gamma\simeq 66$ MeV in
the $e^+e^-\to \pi^+\pi^- h_c$ cross sections~\cite{BESIII:2016adj}. The resonance parameters of
$Y(4220)$ are consistent with those of the resonance observed in the
$e^+e^-\to \omega \chi_{c0}$~\cite{Ablikim:2014qwy}. The newly observed $Y(4220)$ might be a candidate
of $\psi(4S)$ as well. To establish the $4S$ states, more studies are needed.

With the screened potential model we predict that the masses of the $4S$ states are about $4.28$ GeV,
while in the linear potential model their masses are about $4.41$ GeV.
Taking the predicted masses of $\psi(4S)$ with $4412$ MeV and $4281$ GeV
from the linear and screened potential models, respectively,
we calculate the radiative transitions of the $\psi(4S)$ state within these two models.
Our results are listed in Tab.~\ref{EM4s}. It is found that
both the linear and screened potential models give comparable predictions of the decay rates for the $4S$ states
in the magnitude, although the details are different. The decay rates of $4S\to 2 P,3P$
are sizeable, the partial widths for the transitions
$\psi(4S)\to \chi_{cJ}(2P) \gamma$ are about $10\sim 20$ keV, and for the transitions
$\psi(4S)\to \chi_{cJ}(3P) \gamma$ are about $20\sim 80$ keV.
Combined the predicted widths $\Gamma\simeq 78$ MeV from Ref.~\cite{Swanson:2005}, the
branching ratios for the $4S\to 2 P,3P$ transitions are $\mathcal{O}(10^{-4})$.

In coming years, BESIII plan to collect $5\sim 10$ fb$^{-1}$ data samples at
$\psi(4S)$~\cite{XYZ2016}. Using the cross section of
$\sim 4$ nb based on BES measurements~\cite{Bai:2001ct,Mo:2010bw},
we expect to accumulate $(2\sim 4) \times 10^7$
$\psi(4S)$'s. To know the production possibilities of $2P$ and $3P$ states via the
radiative decay of $\psi(4S)$, we estimate the number of
production events from the two-photon cascades $\psi(4S) \to \chi_{cJ}(2P,3P)\gamma\to \psi(1S,2S)
\gamma\gamma$, which has been listed in Tab.~\ref{4sp}.
Unfortunately, it is found that the higher $2P$ and $3P$ states are not able to be produced
via these radiative decay chains at BESIII.

\subsubsection{Higher multipole contributions}

In our calculations, we find that the corrections from the magnetic part
to some radiative transitions of the $S$-, $P$- and $D$-wave states are notable (see Tabs.\ref{EMPW1}--\ref{D-wave}).
For example the magnetic part could give a $10-30\%$ correction
to the radiative partial decay widths of $\Gamma[\chi_{cJ}(1P)\to J/\psi \gamma]$,
$\Gamma[\psi_1(1D)\to \chi_{c1,2}(1P)\gamma]$ and $\Gamma[\psi(3S)\to \chi_{cJ}(1P) \gamma]$.
This large correction is mainly caused by the interferences between the ``extra" electric-dipole
term $E_R$ from the magnetic part and the leading E1 transitions. About
the higher order EM corrections to the radiative transitions, some discussions can be found in the literature\cite{Dudek:2006ej,Dudek:2009kk,Godfrey:1985ei,Grotch:1984gf,Karl:1980wm,
Karl:1975jp,Doncheski:1990kv,Sebastian:1992xq,McClary:1983xw,Rosner:2008ai}.

In experiments, the higher order amplitudes for the transitions $\chi_{c1,2}(1P)\to J/\psi \gamma$
and/or $\psi(2S)\to \chi_{c 1,2}(1P) \gamma$ have been measured in different experiments~\cite{Armstrong:1993fk,Ambrogiani:2001jw,Artuso:2009aa,Ablikim:2004qn,
Oreglia:1981fx,Ablikim:2011da}. Our predictions
with both the linear and screened potential models compared with the data
have been listed in Tab.~\ref{tabM2}. From the table, it is seen that both models give comparable results.
The predicted ratios between the magnetic quadrupole amplitude and
the electric-dipole amplitude, $a_2/a_1$, for the $\chi_{c1,2}(1P)\to J/\psi \gamma$ processes
are in good agreement with the recent measurements from CLEO~\cite{Artuso:2009aa}.
The ratios of $a_2/a_1$ for the $\psi(2S)\to \chi_{c 1,2}(1P) \gamma$ are small.
Their absolute values are comparable to the measurements from CLEO~\cite{Artuso:2009aa}, however,
the sign of $a_2/a_1$ predicted by us seems to be opposite to the measurements.
It should be mentioned that our prediction of
$a_2/a_1$ for the $\psi(2S)\to \chi_{c2}(1P) \gamma$ is
consistent with the previous measurement from BESII~\cite{Ablikim:2004qn}.
More accurate measurements may clarify the sign problem.
The ratios between the ``extra" electric-dipole amplitudes $E_R$
and the $a_1$ are also predicted. It is found that $|\frac{E_R}{a_1}|\simeq |\frac{a_2}{a_1}|$.

Furthermore, we predict the ratios $E_R/a_1$ and $a_2/a_1$ for
some unmeasured processes $\psi_1(1D)\to \chi_{c1,2}(1P)\gamma$ and
$\chi_{cJ}(nP)\to \psi(mS) \gamma$, in which the magnetic
part plays an important role. Our results have been listed in Tab.~\ref{tabM22}.
From the table, it found that most of the ratios are fairly large.
Some ratios can reach to $\sim30\%$. Since the $\psi_1(1D)$
and $\chi_{c2}(2P)$ have been established, and the ratios of $a_2/a_1$ for
$\psi_1(1D)\to \chi_{c1,2}(1P)\gamma$ and $\chi_{c2}(2P)\to J\psi \gamma$
are fairly large, we suggest the experimentalists measure
the ratios of $a_2/a_1$ for these transitions in future experiments.

\section{Summary}\label{sum} %

In this work we calculate the charmonium
spectrum with two models, linear potential model and screened potential model.
We should emphasize that (i) the hyperfine and fine splittings show less model dependence.
The predicted splitting, $m(2 ^3P_{2})-m(2 ^3P_{1})\simeq 90$ MeV,
does not support the $X(3915)$ assigned as
the $\chi_{c0}(2P)$ state. (ii) In the screened potential model, the states $X(4260)$ and $X(4360)$ with $J^{PC}=1^{--}$
may be good candidates of the $\psi(4S)$ and $\psi_1(3D)$, respectively.
(iii) For the newly confirmed $J^{PC}=1^{++}$
states $X(4140)$ and $X(4274)$ by the LHCb,
within the linear potential model the $X(4274)$ might
be identified as the $\chi_{c1}(3P)$ states.
While within the screened potential model, the $X(4140)$ is a good candidate
of $\chi_{c1}(3P)$.

Second, we further evaluate the EM transitions of charmonium states
up to the $4S$ multiplet. It is found that (i) for the EM transitions of
the well-established low-lying charmonium states $J/\psi$, $\psi(2S)$,
$\chi_{cJ}(1P)$, $h_c(1P)$ and $\psi(3770)$, both linear
potential and screened potential models give similar descriptions, which are in
reasonable agreement with the measurements.
(ii) Identifying the newly observed state $X(3823)$ at Belle and BESIII as the $\psi_2(1D)$,
its EM decay properties of are in good agreement with the measurements.
(iii) Assigning the $X(3872)$
resonance as the $\chi_{c1}(2P)$ state, the ratio $\frac{\Gamma[X(3872)\to
\psi(2S) \gamma]}{\Gamma[X(3872)\to J/\psi \gamma]}\simeq 1.3$ predicted by us
is close to the lower limit of the measurements from the BaBar
and LHCb. Thus the $X(3872)$ as the $\chi_{c1}(2P)$ can not be excluded.

Thirdly, we discuss the observations of the missing charmonium states by using radiative
transitions. (i) The large $B\bar{B}$ data sample from Belle II should
let us have chances to establish the missing $\eta_{c2}(1D)$ and $\psi_{3}(1D)$
states in forthcoming experiments.
The $\eta_{c2}(1D)$ state should be produced via the $B\to \eta_{c2}(1D) K$ process
and reconstructed in the $h_c(1P) \gamma$ decay mode with $h_c(1P)\to \eta_c \gamma$ and $\eta_c\to K\bar{K}\pi$.
While the $\psi_{3}(1D)$ state should be produced via the
$B\to \psi_3(1D) K$ process, and reconstructed in the $\chi_{c2}(1P) \gamma$
decay mode with $\chi_{c2}(1P)\to J/\psi \gamma$ and $J/\psi\to \mu^+\mu^-/e^+e^-$.
(ii) If BESIII can accumulate $5\sim 10$ fb$^{-1}$ $\psi(4040)$ data sample
in the coming years, significant numbers of $\chi_{c2}(2P)$ is to be
produced via the radiative decay of $\psi(4040)$, and reconstructed in the $J/\psi \gamma$
decay mode with $J/\psi\to \mu^+\mu^-$.
(iii) Relatively large data samples of $2D$-wave states $\psi_{2}(2D)$ and $\eta_{c2}(2D)$
might be collected at Belle II or LHCb via $B\to \psi_{2}(2D)X$ and
$B\to \eta_{c2}(2D) X$ decays in forthcoming experiments,
the two-photon decay chains $2^3D_2\to \chi_{c1}(1P)\gamma\to
J/\psi \gamma\gamma\to \gamma\gamma \mu^+\mu^-$ ($Br\simeq 1.5\times 10^{-5}$), and
$\eta_{c2}(2D)\to h_{c}(1P)\gamma\to \eta_c(1S)
\gamma\gamma\to K\bar{K}\pi \gamma\gamma$ ($Br\simeq 4.8\times 10^{-5}$) are worth observing.
(iv) The missing $3P$-wave states might be observed at LHCb and Belle II
in the $B\to \chi_{c1,2}(3P)K/h_{c}(3P)K$ decays, and reconstructed in
the $\psi(1S,2S)\gamma$/$\eta_c(2S)\gamma$ decay modes with
$\psi(1S,2S)\to\mu^+\mu^-$/$\eta_c(2S)\to K\bar{K}\pi$.

Finally, we study the corrections of higher EM multipole amplitudes to the EM transitions.
The magnetic part could give about a $10\sim 30\%$ correction
to the radiative partial decay widths of $\Gamma[\chi_{cJ}(1P)\to J/\psi \gamma]$,
$\Gamma[\psi_1(1D)\to \chi_{c1,2}(1P)\gamma]$ and $\Gamma[\psi(3S)\to \chi_{cJ}(1P) \gamma]$.
This large correction is mainly caused by the interferences between the ``extra" electric-dipole
term $E_R$ from the magnetic part and the leading E1 amplitudes. Our predictions for
the normalized magnetic quadrupole amplitude $M_2$ of the $\chi_{c1,2}(1P)\to J/\psi \gamma$ processes
are in good agreement with the recent measurements from CLEO~\cite{Artuso:2009aa}.
About the normalized magnetic quadrupole amplitude of $\psi(2S)\to \chi_{c 1,2}(1P) \gamma$, there
may be a sign difference between our predictions and the measurements.
The normalized ``extra" electric-dipole amplitudes $E_R$ are also predicted.
It is found that $|E_R|\simeq |M_2|$. Furthermore, we find that there are fairly large
magnetic quadrupole amplitudes $M_2$ for the $\chi_{c1,2}(2P,3P)\to \psi(1S,2S) \gamma$ and
$\psi_1(1D)\to \chi_{c1,2}(1P)$ processes. We suggest the experimentalists measure
the higher magnetic quadrupole amplitudes $M_2$ of the $\chi_{2}(2P)\to \psi(1S,2S) \gamma$
and $\psi_1(1D)\to \chi_{c1,2}(1P)$ processes in future experiments.

\section*{  Acknowledgement }

We thank Shi-Lin Zhu and Qiang Zhao for helpful comments and suggestions.
This work is supported, in part, by the National Natural Science
Foundation of China (Grants No. 11075051, No. 11375061, and No.
11405053), and the Hunan Provincial Natural Science Foundation
(Grant No. 13JJ1018).

\begin{table*}[htb]
\caption{Partial widths (keV) of the M1 radiative transitions for
some low-lying $S$-wave charmonium states. LP and SP stand for our results
obtained from the linear potential and screened potential models, respectively. For
comparison,  the predictions from the relativistic quark
model~\cite{Ebert:2003}, NR and GI models~\cite{Swanson:2005} are listed in the table as well. The
experimental average data are taken from the PDG~\cite{PDG}. }\label{EM2}
\begin{tabular}{c|c|cccc|ccccc |cc cc}  \hline\hline
 Initial        & Final  & \multicolumn{4}{|c|} {\underline{$E_{\gamma}$ (MeV)}} & \multicolumn{5}{|c |} {\underline{$\Gamma_{\mathrm{M1}}$ (keV)}} & \underline{$\Gamma_{\mathrm{M1}}$ (keV)}  \\
   state             & state            & \cite{Ebert:2003}& NR\cite{Swanson:2005}&  GI \cite{Swanson:2005}& Ours &  \cite{Ebert:2003}& NR\cite{Swanson:2005}& GI\cite{Swanson:2005} & LP & SP &Exp. \\
\hline
$J/\psi$          &$\eta_{c}(1S)$        & 115 & 116 & 115 & 111    &1.05 & 2.9 & 2.4     &2.39  &2.44    &$1.58\pm0.37$\\
\hline
$\psi(2S)$        &$\eta_{c}(2S)$        & 32~ &48   & 48  & ~47    &0.043& 0.21& 0.17    & 0.19 &0.19   &$0.21\pm0.15$\\
                  &$\eta_{c}(1S)$        & 639 & 639 & 638 & 635    &0.95 & 4.6 & 9.6     & 8.08 &7.80    &$1.24\pm0.29$ \\
$\eta_{c}(2S)$    &$J/\psi$              & 514 & 501 & 501 & 502    &1.53 & 7.9 & 5.6     & 2.64  &2.29    &  \\
\hline
$\psi(3S)$        &$\eta_{c}(3S)$        &     & 29  & 35~ & 30/36    &   & 0.046 & 0.067   & 0.051 & 0.088  &  \\
                  &$\eta_{c}(2S)$        &     & 382 & 436 & 381    &   & 0.61  & 2.6     & 1.65 &1.78     &  \\
                  &$\eta_{c}(1S)$        &     & 922 & 967 & 918    &   & 3.5   & 9.0      & 6.66 &6.76    &  \\
\hline\hline
\end{tabular}
\end{table*}

\begin{table*}[htb]
\caption{Partial widths $\Gamma$ (keV) and branching ratios $Br$ for the radiative transitions (E1 dominant) between the low-lying charmonium states. LP and SP stand for our results
obtained from the linear potential and screened potential models, respectively. For
comparison,  the predictions from the relativistic quark
model~\cite{Ebert:2003}, NR and GI models~\cite{Swanson:2005} and
SNR model~\cite{Li:2009zu} are listed in the table as well. The
experimental average data are taken from the PDG.
$\Gamma_{\mathrm{E1}}$ and $\Gamma_{\mathrm{EM}}$ stands for the E1
and EM transition widths, respectively. }\label{EMPW1}
\begin{tabular}{c|c|c|ccccc|ccc|cccc}  \hline\hline
 Initial        & Final  & \underline{$E_{\gamma}$ (MeV)} & \multicolumn{5}{|c|} {\underline{$\Gamma_{\mathrm{E1}}$ (keV)}}
 & \multicolumn{3}{|c|} {\underline{$\Gamma_{\mathrm{EM}}$ (keV)}}
 & \multicolumn{3}{|c} {\underline{$Br$ (\%)}} \\
   state             & state        &    Ours
 &  \cite{Ebert:2003}& NR/GI~\cite{Swanson:2005}& SNR$_{0/1}$~\cite{Li:2009zu} & LP &SP & LP &SP &Exp.& LP &SP &Exp. \\
\hline
$\psi(2S)$              &$\chi_{c2}(1P)$   &128  & 18.2& 38 / 24   &43/34    & 36&44    &38   &46  &$25.2\pm 2.9$&13.3 &15.7& 9.1$\pm$ 0.3 \\
                        &$\chi_{c1}(1P)$   &171  & 22.9& 54 / 29   &62/36    & 45&48    &42   &45  &$25.5\pm 2.8$&14.7 &15.7&9.6$\pm$ 0.3 \\
                        &$\chi_{c0}(1P)$   &261  & 26.3& 63 / 26   &74/25    & 27&26    &22   &22  &$26.3\pm 2.6$&7.7 &7.7&10.0$\pm$ 0.3 \\
$\eta_{c}(2S)$          &$h_{c}(1P)$       &112  & 41  & 49 / 36   & 146/104 & 49&52    &49   &52  &             &0.43 &0.46&\\
\hline
$\chi_{c2}(1P)$         &      $J/\psi$    & 429 & 327 & 424 / 313 &473/309  & 327&338  &284  &292 &$371\pm 34$ &14.6 &15.0&19.2$\pm$ 0.7\\
$\chi_{c1}(1P)$         &                  & 390 & 265 & 314 / 239 &354/244  & 269&278  &306  &319 &$285\pm14$  &34.8 &36.3&33.9$\pm$ 1.2\\
$\chi_{c0}(1P)$         &                  & 303 & 121 & 152 / 114 &167/117  & 141&146  &172  &179 &$133\pm8$   &1.6 &1.7&1.3$\pm$ 0.1\\
$h_{c}(1P)$             &$\eta_{c}(1S)$    & 499 & 560 & 498 / 352 &764/323  & 361&373  &361  &373 &$357\pm280$ &51.6 &51.0&51.0$\pm$ 6.0\\
\hline
$\psi_1(1D)$            &$\chi_{c2}(1P)$   & 215 & 6.9 & 4.9/3.3   &5.8/4.6  & 5.4&5.7  &7.1 &8.1&$<24.8$     &4.8$\times10^{-2}$ &5.3$\times10^{-2}$& $<9.0\times10^{-2}$\\
                        &$\chi_{c1}(1P)$   & 258 & 135 & 125/77    &150/93   & 115&111  &138  &135 &$81\pm 27$  &0.55 &0.58&0.29$\pm$ 0.06\\
                        &$\chi_{c0}(1P)$   & 346 & 355 & 403/213   &486/197  & 243&232  &272  &261 &$202\pm 42$ &0.99 &0.95&0.73$\pm$ 0.09\\
$\psi_{2}(1D)$          &$\chi_{c2}(1P)$   & 258 & 59  & 64/66     &70/55    & 79 &82   &91   &96  &            &13.3 &14.1&\\
                        &$\chi_{c1}(1P)$   & 299 & 215 & 307/268   &342/208  & 281&291  &285  &296 &            &41.9 &43.5&\\
\hline
\end{tabular}
\end{table*}

\begin{table*}[htb]
\caption{ Partial widths $\Gamma$ (keV) and branching ratios $Br$ for the radiative transitions of the higher $S$-wave states. LP and SP stand for our results
obtained from the linear potential and screened potential models, respectively. For comparison, the predictions from the NR
and GI models~\cite{Swanson:2005} are
listed in the table as well. }\label{EM4s}
\begin{tabular}{c|c|cccc|cccc|cc|ccc}  \hline\hline
 Initial        & Final  & \multicolumn{4}{|c|} {\underline{$E_{\gamma}$ (MeV)}} & \multicolumn{4}{|c|} {\underline{$\Gamma_{\mathrm{E1}}$ (keV)}}
 & \multicolumn{2}{|c} {\underline{$\Gamma_{\mathrm{EM}}$ (keV)}} & \multicolumn{2}{|c} {\underline{$Br$}}  \\
   state             & state            &  NR~\cite{Swanson:2005}&GI~\cite{Swanson:2005}&   LP
   &SP&  NR~\cite{Swanson:2005}&GI~\cite{Swanson:2005}&  LP&SP & LP&SP& LP&SP \\
\hline
$\psi(3S)$      &$\chi_{c2}(2P)$     & 67  &119   &  111&111   & 14    & 48   &65   &79    &67   &82   &$8.4\times 10^{-4}$ & $1.0\times 10^{-3}$  \\
  $80\pm 10$\footnote{Width (MeV) from the PDG~\cite{PDG}.}          &$\chi_{c1}(2P)$     & 113 &145   &  138&138   & 39    & 43   &58   &71    &55   &67   &$6.9\times 10^{-4}$ &$8.4\times 10^{-4}$  \\
                &$\chi_{c0}(2P)$     & 184 &180   &  167&187   & 54    & 22   &21   &31    &19   &27   &$2.4\times 10^{-4}$ & $3.4\times 10^{-4}$ \\
                &$\chi_{c2}(1P)$     & 455 &508   &  455&455   & 0.7   & 13   &0.21 &2.1   &0.25 &2.5  &$3.1\times 10^{-6}$ & $3.1\times 10^{-5}$  \\
                &$\chi_{c1}(1P)$     & 494 &547   &  494&494   &0.53   & 0.85 &4.8  &8.0   &4.0  &6.7  &$5.0\times 10^{-5}$ &$8.4\times 10^{-5}$   \\
                &$\chi_{c0}(1P)$     & 577 &628   &  577&577   &0.27   & 0.63 &9.1  &10.6  &5.9  &6.7  &$7.4\times 10^{-5}$ &$8.4\times 10^{-5}$   \\
$\eta_{c}(3S)$  &$h_{c}(2P)$         & 108 &108   &  108&108   & 105   & 64   &104  &128   &104  &128  &$1.3\times 10^{-3}$ &$1.6\times 10^{-3}$   \\
    80\footnote{Predicted width (MeV) from Ref.~\cite{Swanson:2005}.}   &$h_{c}(1P)$   & 485 &511   &  456&456   & 9.1   & 28   &0.045&1.4   &0.045&1.4  &$5.6\times 10^{-7}$ &$1.8\times 10^{-5}$\\
$\psi(4S)$      &$\chi_{c2}(1P)$     & 775 &804   &  773&664   & 0.61  &5.2   &0.13 &0.66  &0.17 &0.84 &$2.2\times 10^{-6}$ & $1.1\times 10^{-5}$ \\
    78$^b$      &$\chi_{c1}(1P)$     & 811 &841   &  809&701   & 0.41  &0.53  &3.8  &3.9   &2.9  &3.0 &$3.7\times 10^{-5}$&$3.8\times 10^{-5}$ \\
                &$\chi_{c0}(1P)$     & 887 &915   &  884&778   & 0.18  &0.13  &7.5  &6.2   &3.7  &2.7  &$4.7\times 10^{-5}$ &$4.2\times 10^{-5}$  \\
                &$\chi_{c2}(2P)$     & 421 &446   &  458&339   & 0.62  &15    &11   &4.7   &13   &5.3  &$1.7\times 10^{-4}$ &$6.8\times 10^{-5}$\\
                &$\chi_{c1}(2P)$     & 423 &469   &  482&364   & 0.49  &0.92  &24   &12    &20   &11   &$2.6\times 10^{-4}$ &$1.4\times 10^{-4}$\\
                &$\chi_{c0}(2P)$     & 527 &502   &  510&411   & 0.24  &0.39  &17   &12    &12   &8.7  &$1.5\times 10^{-4}$ &$1.1\times 10^{-4}$\\
                &$\chi_{c2}(3P)$     & 97  &112   &  101&69    &   68  &66    &80   &39    &82   &40   &$1.1\times 10^{-3}$ &$5.1\times 10^{-4}$\\
                &$\chi_{c1}(3P)$     & 142 &131   &  126&88    &  126  &54    &74   &38    &71   &37   &$9.1\times 10^{-4}$ &$4.7\times 10^{-4}$\\
                &$\chi_{c0}(3P)$     & 208 &155   &  178&133   &0.003  & 25   &40   &23    &36   &21   &$4.6\times 10^{-4}$ &$2.7\times 10^{-4}$\\
$\eta_c(4S)$    &$h_{c}(1P)$         & 782 &808   &  778&675   &  5.2  &9.6   &0.29 &0.63  &0.29 &0.63 &$4.8\times 10^{-6}$ &$1.0\times 10^{-5}$ \\
    61$^b$          &$h_{c}(2P)$         & 427 &444   &  461&348   & 10.1  &31.3  &20   &7.9  &20 &7.9 &$3.3\times 10^{-4}$ &$1.3\times 10^{-4}$ \\
                &$h_{c}(3P)$         & 104 &106   &  142&70    &  159  &101   &102  &70    &102  &70   &$1.7\times 10^{-3}$ &$1.1\times 10^{-3}$ \\
\hline

\end{tabular}
\end{table*}

\begin{table*}[htb]
\caption{ Partial widths $\Gamma$ (keV) and branching ratios $Br$ for the radiative transitions (E1 dominant) of the higher $2P$ and $3P$ states. LP and SP stand for our results
obtained from the linear potential and screened potential models, respectively. For comparison, the predictions from the NR
and GI models~\cite{Swanson:2005} and SNR model~\cite{Li:2009zu} are
listed in the table as well. }\label{EMPA}
\begin{tabular}{c|c|ccc|cccc|cc|cc}  \hline\hline
 Initial        & Final  & \multicolumn{3}{|c|} {\underline{$E_{\gamma}$ (MeV)}} & \multicolumn{4}{|c|} {\underline{$\Gamma_{\mathrm{E1}}$ (keV)}}
 & \multicolumn{2}{|c} {\underline{$\Gamma_{\mathrm{EM}}$ (keV)}} & \multicolumn{2}{|c} {\underline{$Br$}} \\
   state             & state            &  NR/GI~\cite{Swanson:2005}&  SNR~\cite{Li:2009zu}& LP/SP
   &  NR/GI~\cite{Swanson:2005}& SNR$_{0/1}$~\cite{Li:2009zu} & LP&SP & LP&SP & LP&SP\\
\hline
$\chi_{c2}(2P)$    &$\psi_{3}(1D)$      & 163 /128 &     &96/96    & 88    / 29   &        &20   &24    &20  &24   &$8.3\times 10^{-4}$ &$1.0\times 10^{-3}$ \\
     24$\pm$6\footnote{Width (MeV) from the PDG~\cite{PDG}.}            &$\psi_{2}(1D)$      & 168 /139 &     & 103     & 17    / 5.6  &        & 3.3 &4.1   &3.2 &4.0  &$1.3\times 10^{-4}$ &$1.7\times 10^{-4}$ \\
                   &$\psi_1(1D)$        & 197 /204 &     & 146     & 1.9   / 1.0  &        &0.47 &0.62  &0.36&0.46 &$1.5\times 10^{-5}$ &$1.9\times 10^{-5}$ \\
                   &$\psi(2S)$          & 276 /282 & 235 & 234     & 304   / 207  &225/100 & 146 &163   &135 &150  &$5.6\times 10^{-3}$ &$6.3\times 10^{-3}$ \\
                   &$J/\psi$            & 779 /784 & 744 & 742     & 81    / 53   &101/109 & 118 &119   &93  &93   &$3.9\times 10^{-3}$ &$3.9\times 10^{-3}$ \\
$\chi_{c1}(2P)$    &$\psi_{2}(1D)$      & 123 /113 &     &76/76    & 35    / 18   &        &2.8  &3.4   &2.9 &3.5  &$1.8\times 10^{-5}$ &$2.1\times 10^{-5}$ \\
      165\footnote{Predicted width (MeV) from Ref.~\cite{Swanson:2005}.}          &$\psi_1(1D)$        & 152 /179 &     &120/120  & 22    / 21   &        &8.6  &10.8  &7.9 &9.8  &$4.9\times 10^{-5}$ &$5.9\times 10^{-5}$ \\
                   &$\psi(2S)$          & 232 /258 & 182 &208/208  & 183   / 183  & 103/60 &129  &145   &139 &155  &$8.4\times 10^{-4}$ &$9.4\times 10^{-4}$  \\
                   &$J/\psi$            & 741 /763 & 697 &720/720  & 71    / 14   & 83/45  &64   &68    &81  &88   &$4.9\times 10^{-4}$ &$5.3\times 10^{-4}$ \\
$\chi_{c0}(2P)$    &$\psi_1(1D)$        & 81  /143 &     &90 /69   & 13    / 51   &        &21   &12    &20  &12   &$6.7\times 10^{-4}$ &$4.0\times 10^{-4}$ \\
      $30^b$           &$\psi(2S)$          & 162 /223 & 152 &179/159  & 64    / 135  & 61/44  &108  &89    &121 &99   &$4.0\times 10^{-3}$ &$3.3\times 10^{-3}$\\
                   &$J/\psi$            & 681 /733 & 672 &695/678  & 56    / 1.3  & 74/9.3 &4.0  &1.5   &6.1 &2.3  &$2.0\times 10^{-4}$ &$7.7\times 10^{-5}$ \\
$h_{c}(2P)$        &$\eta_{c2}(1D)$     & 133 /117 &     &100/100  & 60    / 27   &        &25   &25    &25  &25   &$2.9\times 10^{-4}$ &$2.9\times 10^{-4}$ \\
      $87^b$           &$\eta_{c}(2S)$      & 285 /305 & 261 &252/252  & 280   / 218  & 309/108&160  &176   &160 &176  &$1.8\times 10^{-3}$ &$2.0\times 10^{-3}$ \\
                   &$\eta_{c}(1S)$      & 839 /856 & 818 &808/808  & 140   / 85   & 134/250&135  &134   &135 &134  &$1.6\times 10^{-3}$ &$1.6\times 10^{-3}$ \\
\hline
$\chi_{c2}(3P)$    &$\psi_{3}(2D)$      & 147 /118 &     &136/98   & 148   / 51   &        &116  &64    &121 &66   &$1.8\times 10^{-3}$ &$1.0\times 10^{-3}$ \\
       $66^b$            &$\psi_{2}(2D)$      & 156 /127 &     &143/101  & 31    / 10   &        &18   &10    &18  &10   &$2.7\times 10^{-4}$ &$1.5\times 10^{-4}$ \\
                   &$\psi_1(2D)$        & 155 /141 &     &117/20   & 2.1   / 0.77 &        &0.55 &0.004 &0.44&0.004&$6.7\times 10^{-6}$ &$6.0\times 10^{-8}$ \\
                   &$\psi_{3}(1D)$      & 481 /461 &     &453/364  & 0.049 / 6.8  &        &15   &10    &17  &12   &$1.1\times 10^{-4}$ &$1.8\times 10^{-4}$ \\
                   &$\psi_{2}(1D)$      & 486 /470 &     &459/370  & 0.01  / 0.13 &        &4.6  &2.5   &4.6  &2.4  &$7.0\times 10^{-5}$ &$3.6\times 10^{-5}$ \\
                   &$\psi_1(1D)$        & 512 /530 &     &495/411  & 0.00  / 0.00 &        &1.9  &1.0   &1.5 &0.79  &$2.2\times 10^{-5}$ &$1.2\times 10^{-5}$ \\
                   &$\psi(3S)$          & 268 /231 &     &261/168  & 509   / 199  &        &306  &121   &281 &114  &$4.3\times 10^{-3}$ &$1.7\times 10^{-3}$ \\
                   &$\psi(2S)$          & 585 /602 &     &574/492  & 55    / 30   &        &116  &90    &97  &76   &$1.5\times 10^{-3}$ &$1.2\times 10^{-3}$ \\
                   &$J/\psi$            & 1048/1063&     &1042/967 & 34    / 19   &        &83   &69    &61  &51   &$9.2\times 10^{-4}$ &$7.7\times 10^{-4}$ \\
$\chi_{c1}(3P)$    &$\psi_{2}(2D)$      & 112 /108 &     &117/82   & 58    / 35   &        &22   &11    &23  &11   &$5.9\times 10^{-4}$ &$2.8\times 10^{-4}$ \\
        $39^b$           &$\psi_1(2D)$        & 111 /121 &     &92 /1    & 19    / 15   &        &8.6  &0     &8.1 &0    &$2.1\times 10^{-4}$ &0 \\
                   &$\psi_{2}(1D)$      & 445 /452 &     &436/353  & 0.035 / 4.6  &        &0.13 &0.09  &0.12&0.09 &$3.1\times 10^{-6}$ &$2.3\times 10^{-6}$  \\
                   &$\psi_1(1D)$        & 472 /512 &     &476/394  & 0.014 / 0.39 &        &4.4  &2.7   &3.2 &2.0  &$6.1\times 10^{-5}$ &$4.1\times 10^{-5}$ \\
                   &$\psi(3S)$          & 225 /212 &     &237/149  & 303   / 181  &        &305  &111   &331 &117  &$8.5\times 10^{-3}$ &$3.0\times 10^{-3}$ \\
                   &$\psi(2S)$          & 545 /585 &     &556/475  & 45    / 8.9  &        &78   &63    &94  &74   &$2.4\times 10^{-3}$ &$1.9\times 10^{-3}$ \\
                   &$J/\psi$            & 1013/1048&     &1023/952 & 31    / 2.2  &        &36   &33    &50  &45   &$1.3\times 10^{-3}$ &$1.2\times 10^{-3}$ \\
$\chi_{c0}(3P)$    &$\psi_1(2D)$        & 43  /97  &     &39 /45   & 4.4   / 35   &        &3.8  &9.3   &3.8 &9.1  &$7.5\times 10^{-5}$ &$1.8\times 10^{-4}$  \\
       $51^b$            &$\psi_1(1D)$        & 410 /490 &     &427/352  & 0.037 / 9.7  &        &0.31 &0.44  &0.27&0.39 &$5.3\times 10^{-6}$ &$7.6\times 10^{-6}$\\
                   &$\psi(3S)$          &159  /188 &     &186/105  & 109   / 145  &        &214  &56    &241 &61   &$4.7\times 10^{-3}$ &$1.2\times 10^{-3}$\\
                   &$\psi(2S)$          &484  /563 &     &509/434  & 32    / 0.045&        &13   &6.9   &17  &9.1  &$3.3\times 10^{-4}$ &$1.8\times 10^{-4}$\\
                   &$J/\psi$            &960  /1029&     &981/916  & 27    / 1.5  &        &0.14 &0.08  &0.24&0.13  &$4.7\times 10^{-6}$ &$2.5\times 10^{-6}$\\
$h_{c}(3P)$        &$\eta_{c2}(2D)$     &119  /109 &     &120/84   & 99    / 48   &        &93   &47    &93&47    &$1.2\times 10^{-3}$ &$6.3\times 10^{-4}$\\
       $75^b$            &$\eta_{c2}(1D)$     &453  /454 &     &453/370   & 0.16 / 5.7  &        &15   &8.7    &15&8.7   &$2.0\times 10^{-4}$ &$1.2\times 10^{-4}$\\
                   &$\eta_{c}(3S)$      &229  /246 &     &238/185   & 276  / 208  &        &237  &146   &237&146   &$3.2\times 10^{-3}$ &$1.9\times 10^{-3}$\\
                   &$\eta_{c}(2S)$      &593  /627 &     &602/517   & 75   / 43   &        &124  &96    &124&96   &$1.7\times 10^{-3}$ &$1.3\times 10^{-3}$\\
                   &$\eta_{c}(1S)$      &1103 /1131&     &1104/1035 & 72   / 38   &        &90   &77    &90&77 &$1.2\times 10^{-4}$ &$1.0\times 10^{-3}$\\
\hline\hline

\end{tabular}
\end{table*}


\begin{table*}[htb]
\caption{Partial widths $\Gamma$ (keV) and branching ratios $Br$ for the radiative transitions (E1 dominant) of
the higher $D$-wave states. LP and SP stand for our results
obtained from the linear potential and screened potential models, respectively.
For comparison, the predictions from the relativistic quark model~\cite{Ebert:2003},
NR and GI models~\cite{Swanson:2005}, and SNR model~\cite{Li:2009zu}
are listed in the table as well. }\label{D-wave}
\begin{tabular}{c|c|cccc|ccccc|cc|cccc}  \hline\hline
 Initial      & Final  & \multicolumn{4}{|c|} {\underline{$E_{\gamma}$ (MeV)}} & \multicolumn{5}{|c|} {\underline{$\Gamma_{\mathrm{E1}}$ (keV)}}
 & \multicolumn{2}{|c|} {\underline{$\Gamma_{\mathrm{EM}}$ (keV)}}& \multicolumn{2}{|c} {\underline{$Br$ }}  \\
   state             & state            & \cite{Ebert:2003}& NR/GI~\cite{Swanson:2005}&  SNR~\cite{Li:2009zu}
   &  LP/SP& \cite{Ebert:2003}& NR/GI~\cite{Swanson:2005} & SNR$_{0/1}$~\cite{Li:2009zu} & LP&SP&LP&SP&LP&SP \\
\hline
$\psi_{3}(1D)$         &$\chi_{c2}(1P)$   & 250      & 242/282 & 236     &264/264       & 156 & 272 /296 & 284/223  &377 &393  & 350 &364 &12\%&12\%   \\
$\eta_{c2}(1D)$        &$h_{c}(1P)$       & 275      & 264/307 & 260     &284/284       & 245 & 339 /344 & 575/375  &362 &376  & 362 &376 &72\%&75\%  \\
\hline
$\psi_{3}(2D)$         &$\chi_{c2}(1P)$   &          & 566/609 &         &571/518       &     &  29 / 16 &          &83  &78   & 72  &67  &$4.9\times 10^{-4}$&$4.5\times 10^{-4}$  \\
    148\footnote{Predicted width (MeV) from Ref.~\cite{Swanson:2005}.}                   &$\chi_{c2}(2P)$   &          & 190/231 &         &238/181       &     & 239 /272 &          &457 &256  & 427 &243 &$2.9\times 10^{-3}$& $1.6\times 10^{-3}$ \\
$\psi_{2}(2D)$        &$\chi_{c2}(1P)$   &          & 558/602 &         &564/516       &     & 7.1 /0.62&          &16  &16   &  20 &20  &$1.7\times 10^{-4}$& $2.2\times 10^{-4}$ \\
    92$^{a}$          &$\chi_{c1}(1P)$   &          & 597/640 &         &603/554       &     &  26 / 23 &          &64  &64   & 68  &68  &$7.4\times 10^{-4}$& $7.4\times 10^{-4}$ \\
                       &$\chi_{c2}(2P)$   &          & 182/223 &         &231/178       &     & 52  / 65 &          &101 &57   & 115 &64  &$1.3\times 10^{-3}$& $7.0\times 10^{-4}$  \\
                       &$\chi_{c1}(2P)$   &          & 226/247 &         &222/204       &     & 298 /225 &          &220 &186  & 223 &188 &$2.4\times 10^{-3}$&$2.0\times 10^{-3}$\\
$\psi_1(2D)$           &$\chi_{c2}(1P)$   &          & 559/590 &         & 587          &     &0.79/0.027&          &16  &16   &17   &20  &$2.3\times 10^{-4}$& $2.7\times 10^{-4}$ \\
    74$^{a}$           &$\chi_{c1}(1P)$   &          & 598/628 &         & 625          &     & 14 / 3.4 &          &25  &42   &37   &63  &$5.0\times 10^{-4}$&$8.5\times 10^{-4}$  \\
                       &$\chi_{c0}(1P)$   &          & 677/707 &         & 704          &     & 27 / 35  &          &120 &149  &150  &189 &$2.0\times 10^{-3}$ &$2.6\times 10^{-3}$   \\
                       &$\chi_{c2}(2P)$   &          & 183/210 &         & 256          &     & 5.9/ 6.3 &          &18  &21   &24   &29  &$3.2\times 10^{-4}$ &$3.9\times 10^{-4}$   \\
                       &$\chi_{c1}(2P)$   &          & 227/234 &         &281/281       &     & 168 /114 &          &253 &280  & 309 &347 &$4.2\times 10^{-3}$ &$4.7\times 10^{-3}$  \\
                       &$\chi_{c0}(2P)$   &          & 296/269 &         &312/329       &     & 483 /191 &          &299 &321  & 332 &360 &$4.5\times 10^{-3}$ & $4.9\times 10^{-3}$  \\
$\eta_{c2}(2D)$        &$h_{c}(1P)$       &          & 585/634 &         &590/542       &     &  40 / 25 &          &96 &92  & 96 &92 &$1.3\times 10^{-3}$ & $1.2\times 10^{-3}$   \\
 111$^{a}$             &$h_{c}(2 P)$      &          & 218/244 &         &256/203       &     & 336 /296 &          &438 &271  & 438 &271 &$3.9\times 10^{-3}$ &$2.4\times 10^{-3}$  \\
\hline\hline
\end{tabular}
\end{table*}

\begin{table*}[htb]
\begin{center}
\caption{ Three-photon decay chains of $2^3P_2$. The branching fractions are $Br_1=Br[2^3P_2\to 1^3D_J \gamma]$, $Br_2=Br[1^3D_J \to 1^3P_J\gamma]$, $Br_3=Br[1^3P_J\to J/\psi \gamma]$, and $Br=Br_1\times Br_2\times Br_3$ is the combined branching fractions of the chain. The branching fractions are predicted with the linear potential model.}\label{2p2chain}
\begin{tabular}{c|c|c|c|ccccc}
\hline\hline
  Decay chain    & $Br_1$ & $Br_2$&$Br_3$ &$Br$\\
 \hline
    $ 2^3P_2\to 1^3D_1\to 1^3P_0\to J/\psi$   &$1.5\times 10^{-5}$ &$9.9\times 10^{-3}$  &1.6\%  &$2.4\times 10^{-9}$    \\
    $ 2^3P_2\to 1^3D_1\to 1^3P_1\to J/\psi $   &$1.5\times 10^{-5}$ &$5.5\times 10^{-3}$  &34.8\%  & $2.9\times 10^{-8}$     \\
    $ 2^3P_2\to 1^3D_1\to 1^3P_2\to J/\psi $   &$1.5\times 10^{-5}$ &$4.8\times 10^{-4}$  &14.6\%  &$1.0\times 10^{-8}$     \\
    $ 2^3P_2\to 1^3D_2\to 1^3P_1\to J/\psi $   &$1.3\times 10^{-4}$      &42\%  &34.8\%  &$1.9\times 10^{-5}$      \\
    $ 2^3P_2\to 1^3D_2\to 1^3P_2\to J/\psi $   &$1.3\times 10^{-4}$      &13\%  &14.6\%  &$2.5\times 10^{-6}$     \\
    $ 2^3P_2\to 1^3D_3\to 1^3P_2\to J/\psi $   &$8.3\times 10^{-4}$      &12\%  &14.6\%  &$1.4\times 10^{-5}$      \\
\hline\hline
\end{tabular}
\end{center}
\end{table*}

\begin{table*}[htb]
\begin{center}
\caption{ Two-photon decay chains of $3^3S_1$. The branching fractions are $Br_1=Br[3^3S_1\to 2^3P_J \gamma]$, $Br_2=Br[2^3P_J\to 2^3S_{1}\gamma,J/\psi \gamma]$, $Br_3=Br[2^3S_1,J/\psi\to \mu^+\mu^-]$ (obtained from PDG~\cite{PDG}), and $Br=Br_1\times Br_2\times Br_3$ is
the combined branching fractions of the chain. The theoretical branching fractions are predicted with the linear potential model.
The estimated events are based on producing of $5\times 10^7$ $\psi(3S)$ events at BESIII in coming years as described in the text.}\label{3stop}
\begin{tabular}{c|c|c|c|c|cccccc}
\hline\hline
  Decay chain    & $Br_1(10^{-4})$ & $Br_2(10^{-4})$&$Br_3(\%)$& $Br(10^{-8})$ & Events\\
 \hline
    $3^3S_1\to 2^3P_2\to 2^3S_1\to \mu^+\mu^- $       &  $8.4$  &$56$ & 0.79     &$3.7$  & 2 \\
    $3^3S_1\to 2^3P_1\to 2^3S_1\to \mu^+\mu^- $       &  $6.9$  & $8.4$ &0.79     &$0.46$& 0.2   \\
    $3^3S_1\to 2^3P_0\to 2^3S_1\to \mu^+\mu^- $       & $2.4$  & $40$   & 0.79    &$0.76$& 0.4   \\
    $3^3S_1\to 2^3P_2\to J/\psi\to \mu^+\mu^- $       &  $8.4$  &$39$ & 5.9       &$19$  & 10 \\
    $3^3S_1\to 2^3P_1\to J/\psi\to \mu^+\mu^- $       &  $6.9$  & $4.9$ &5.9      &$2.0$ & 1  \\
    $3^3S_1\to 2^3P_0\to J/\psi\to \mu^+\mu^- $       & $2.4$  & $2.0$   & 5.9    &$0.28$& 0.1   \\
\hline\hline
\end{tabular}
\end{center}
\end{table*}

\begin{table*}[htb]
\begin{center}
\caption{ Two-photon decay chains of $2D$-wave states. The branching fractions are $Br_1=Br[2D \to nP \gamma]$, $Br_2=Br[nP \to 2^3S_1 \gamma, 1S \gamma ]$, $Br_3=Br[2^3S_1,J/\psi\to \mu^+\mu^-]$ (obtained from PDG~\cite{PDG}) or $Br_3=Br[\eta_c(1S)\to K\bar{K}\pi]$ (obtained from PDG~\cite{PDG}), and $Br=Br_1\times Br_2\times Br_3$ is the combined branching fractions of the chain. The theoretical branching fractions are predicted with the linear potential model. The estimated events are based on producing of $2.4\times 10^7$ $\psi(4160)$'s at BESIII as described in the text.}\label{2d2p}
\begin{tabular}{c|c|c|c|cccc}
\hline\hline
  Decay chain  & $Br_1$($10^{-3}$) & $Br_2$($10^{-4}$)&$Br_3$($\%$)~\cite{PDG}&$Br$($10^{-7}$) & Events\\
 \hline
    $2^3D_1\to 2^3P_2\to 2^3S_1\to \mu^+\mu^-$   & $0.32$    &$56$ &0.79  &$0.14$  &0.3  \\
    $2^3D_1\to 2^3P_1\to 2^3S_1\to \mu^+\mu^-$   & $4.2$    &$8.4$ &0.79   &$0.28$ &0.6    \\
    $2^3D_1\to 2^3P_0\to 2^3S_1\to \mu^+\mu^-$   & $4.5$    &$40$ &0.79   &$1.4$   &3  \\
    $2^3D_1\to 2^3P_2\to J/\psi\to \mu^+\mu^-$   & $0.32$    &$39$ &5.9   &$0.74$  &2   \\
    $2^3D_1\to 2^3P_1\to J/\psi\to \mu^+\mu^-$   & $4.2$    &$4.9$ &5.9   &$1.2$   &3    \\
    $2^3D_1\to 2^3P_0\to J/\psi\to \mu^+\mu^-$   & $4.5$    &$2.0$ &5.9   &$0.53$  &1     \\
    $2^3D_1\to 1^3P_2\to J/\psi\to \mu^+\mu^-$   & $0.23$    &$1460$ &5.9   &$19.8$&47    \\
    $2^3D_1\to 1^3P_1\to J/\psi\to \mu^+\mu^-$   & $0.50$    &$3480$ &5.9   &$102$ &244     \\
    $2^3D_1\to 1^3P_0\to J/\psi\to \mu^+\mu^-$   & $2.0$     &$160$ &5.9   &$18.8$ &45     \\
\hline
    $2^3D_2\to 2^3P_2\to 2^3S_1\to \mu^+\mu^-$   & $1.3$    &$56$ &0.79   &$0.57$ & $\cdot\cdot\cdot$   \\
    $2^3D_2\to 2^3P_1\to 2^3S_1\to \mu^+\mu^-$   & $2.4$    &$8.4$ &0.79   &$0.16$& $\cdot\cdot\cdot$      \\
    $2^3D_2\to 2^3P_2\to J/\psi\to \mu^+\mu^-$   & $1.3$    &$39$ &5.9   &$3.0$   & $\cdot\cdot\cdot$  \\
    $2^3D_2\to 2^3P_1\to J/\psi\to \mu^+\mu^-$   & $2.4$    &$4.9$ &5.9   &$6.9$ &  $\cdot\cdot\cdot$   \\
    $2^3D_2\to 1^3P_2\to J/\psi\to \mu^+\mu^-$   & $0.17$    &$1460$ &5.9   &$14.6$ & $\cdot\cdot\cdot$    \\
    $2^3D_2\to 1^3P_1\to J/\psi\to \mu^+\mu^-$   & $0.74$    &$3480$ &5.9   &$152$ & $\cdot\cdot\cdot$    \\

\hline
    $2^3D_3\to 2^3P_2\to 2^3S_1\to \mu^+\mu^-$   & $2.9$    &$56$ &0.79   &$1.3$ & $\cdot\cdot\cdot$    \\
    $2^3D_3\to 2^3P_2\to J/\psi\to \mu^+\mu^-$   & $2.9$    &$39$ &5.9   &$6.7$  & $\cdot\cdot\cdot$   \\
    $2^3D_3\to 1^3P_2\to J/\psi\to \mu^+\mu^-$   & $0.49$    &$1460$ &5.9   &$42$ & $\cdot\cdot\cdot$    \\
    $2^1D_2\to 2^1P_1\to \eta_c(1S)\to K\bar{K}\pi$   & $3.9$    &$16$ &7.3  &$4.6$& $\cdot\cdot\cdot$     \\
    $2^1D_2\to 1^1P_1\to \eta_c(1S)\to K\bar{K}\pi$   & $1.3$ &$5100$ &7.3  &$483$ & $\cdot\cdot\cdot$    \\
 \hline
\hline
\end{tabular}
\end{center}
\end{table*}

\begin{table*}[htb]
\begin{center}
\caption{ Two-photon decay chains of $4^3S_1$. The branching fractions are $Br_1=Br[4^3S_1\to 2^3P_J \gamma,3^3P_J \gamma]$, $Br_2=Br[n^3P_J\to m^3S_{1}\gamma]$, $Br_3=Br[2^3S_1,J/\psi\to \mu^+\mu^-]$ (obtained from PDG~\cite{PDG}), and $Br=Br_1\times Br_2\times Br_3$ is
the combined branching fractions of the chain. The theoretical branching fractions are predicted with the linear potential model. The estimated events are based on producing of $2\times 10^7$ $\psi(4S)$'s at BESIII in coming years as described in the text.}\label{4sp}
\begin{tabular}{c|c|c|c|c|ccccccc}
\hline\hline
  Decay chain   &$Br_1 $$(10^{-4})$ & $Br_2$$(10^{-4})$&$Br_3$$(\%)$~\cite{PDG}& $Br$$(10^{-9})$ & Events\\
 \hline
    $4^3S_1\to 2^3P_2\to 2^3S_1\to \mu^+\mu^- $   &$1.7$ &$56$ &0.79  & $7.5$  &  0.15          \\
    $4^3S_1\to 2^3P_1\to 2^3S_1\to \mu^+\mu^- $   &$2.6$ &$8.4$ &0.79  &$1.7$  &  0.03           \\
    $4^3S_1\to 2^3P_0\to 2^3S_1\to \mu^+\mu^- $   &$1.5$ &$40$ &0.79  & $4.7$  &  0.09           \\
    $4^3S_1\to 2^3P_2\to J/\psi\to \mu^+\mu^- $   &$1.7$ &$39$ &5.9  &$39$    &   0.78         \\
    $4^3S_1\to 2^3P_1\to J/\psi\to \mu^+\mu^- $   &$2.6$ &$4.9$ &5.9  &$7.5$ &    0.15          \\
    $4^3S_1\to 2^3P_0\to J/\psi\to \mu^+\mu^- $   &$1.5$ &$2.0$ &5.9  &$1.8$ &    0.03          \\
 \hline
    $4^3S_1\to 3^3P_2\to 2^3S_1\to \mu^+\mu^- $   &$11 $ &$15$ &0.79  & $13$ &    0.26         \\
    $4^3S_1\to 3^3P_1\to 2^3S_1\to \mu^+\mu^- $   &$9.1$ &$24$ &0.79  &$17$  &    0.34         \\
    $4^3S_1\to 3^3P_0\to 2^3S_1\to \mu^+\mu^- $   &$4.6$ &$33$ &0.79  & $12$ &    0.24          \\
    $4^3S_1\to 3^3P_2\to J/\psi\to \mu^+\mu^- $   &$1.7$ &$9.2$ &5.9  &$9.2$ &    0.18          \\
    $4^3S_1\to 3^3P_1\to J/\psi\to \mu^+\mu^- $   &$2.6$ &$13$ &5.9   &$20$  &    0.40         \\
    $4^3S_1\to 3^3P_0\to J/\psi\to \mu^+\mu^- $   &$1.5$ &$0.047$ &5.9  &$0.041$ &0              \\
\hline\hline
\end{tabular}
\end{center}
\end{table*}

\begin{table*}[htb]
\begin{center}
\caption{The predicted ratios between the magnetic quadrupole amplitude $a_2$ and the electric-dipole amplitude $a_1$ compared with the data. The predicted ratios between the ``extra" electric-dipole $E_R$ and electric-dipole $a_1$ are also listed. LP and SP stand for our results
obtained from the linear potential and screened potential models, respectively.} \label{tabM2}
\begin{tabular}{c|ccccccccc}
\hline\hline
 process &$\frac{E_R}{a_1}$ &$\frac{E_R}{a_1}$ & $\frac{a_2}{a_1}$   &$\frac{a_2}{a_1}$   & $\frac{a_2}{a_1}$   &$\frac{a_2}{a_1}$ &$\frac{a_2}{a_1}$
 & $\frac{a_2}{a_1}$  & $\frac{a_2}{a_1}$    \\
                                      &  LP & SP                   &  SP  & LP & Lat.~\cite{Dudek:2009kk} &CLEO~\cite{Artuso:2009aa}  &BESII~\cite{Ablikim:2004qn}&Crystal Ball~\cite{Oreglia:1981fx} & BESIII~\cite{Ablikim:2011da} \\
 \hline
$\chi_{c1}(1P)\to J/\psi \gamma$      &$+0.062$ & $+0.065$& $-0.065$       & $-0.062$ & $-0.09(7)$&$-0.0626(87)$ &&$-0.002^{+0.008}_{-0.020}$ & \\
$\chi_{c2}(1P)\to J/\psi \gamma$      &$-0.078$ & $-0.082$& $-0.110$       & $-0.105$ &$-0.39(7)$              &$-0.093(19)$ & &$-0.333^{+0.116}_{-0.292}$ &\\
$\psi(2S)\to \chi_{c1}(1P) \gamma$    &$-0.030$ & $+0.031$& $-0.031$        & $-0.030$ &              &0.0276(96) & &$0.077^{+0.050}_{-0.045}$ &\\
$\psi(2S)\to \chi_{c2}(1P) \gamma$    &$+0.021$ & $+0.022$& $-0.030$        & $-0.028$ &              &0.010(16) &$-0.051^{+0.054}_{-0.036}$ &$0.132^{+0.098}_{-0.075}$ &$0.046(23)$\\
\hline\hline
\end{tabular}
\end{center}
\end{table*}

\begin{table*}[htb]
\begin{center}
\caption{The predicted ratios $\frac{a_2}{a_1}$ and $\frac{E_R}{a_1}$ with the linear potential (LP) and screened potential (SP) models. } \label{tabM22}
\begin{tabular}{c|ccccccccc}
\hline\hline
 process &$\frac{E_R}{a_1}$(LP) &$\frac{E_R}{a_1}$(SP) & $\frac{a_2}{a_1}$ (SP) &$\frac{a_2}{a_1}$ (LP)   \\
 \hline
$\psi_1(1D)\to \chi_{c1}(1P) \gamma$  &$+0.088$& $+0.092$&  $+0.041$  & $+0.040$  \\
$\psi_1(1D)\to \chi_{c2}(1P) \gamma$  &$+0.214$& $+0.224$&  $+0.074$  & $+0.066$  \\
$\chi_{c1}(2P)\to J/\psi \gamma$      &$+0.108$& $+0.113$&  $-0.113$  & $-0.108$  \\
$\chi_{c2}(2P)\to J/\psi \gamma$      &$-0.143$& $-0.151$&  $-0.203$  & $-0.192$  \\
$\chi_{c1}(2P)\to \psi(2S) \gamma$    &$+0.034$& $+0.036$&  $-0.036$  & $-0.034$  \\
$\chi_{c2}(2P)\to \psi(2S) \gamma$    &$-0.041$& $-0.043$&  $-0.058$  & $-0.055$  \\
$\chi_{c1}(3P)\to J/\psi \gamma$      &$+0.147$& $+0.144$&  $-0.144$  & $-0.147$ \\
$\chi_{c2}(3P)\to J/\psi \gamma$      &$-0.213$& $-0.207$&  $-0.277$  & $-0.286$\\
$\chi_{c1}(3P)\to \psi(2S) \gamma$    &$+0.086$& $+0.078$&  $-0.078$  &$-0.086$ \\
$\chi_{c2}(3P)\to \psi(2S) \gamma$    &$-0.107$& $-0.096$&  $-0.128$  &$-0.144$ \\
$\chi_{c1}(3P)\to \psi(3S) \gamma$    &$+0.038$& $+0.026$&  $-0.026$  & $-0.038$  \\
$\chi_{c2}(3P)\to \psi(3S) \gamma$    &$-0.046$& $-0.031$&  $-0.041$  & $-0.062$  \\
\hline\hline
\end{tabular}
\end{center}
\end{table*}

\end{document}